\documentclass[aps,prl,twocolumn,showpacs,superscriptaddress,groupaddress,floatfix,graphics,graphicx]{revtex4-1}
\usepackage[latin9]{inputenc}
\usepackage{color}
\usepackage{amsmath}
\usepackage{amssymb}
\usepackage{graphicx}
\usepackage{bbold}
\usepackage{bm}
\usepackage[unicode=true,pdfusetitle,
 bookmarks=true,bookmarksnumbered=false,bookmarksopen=false,
 breaklinks=false,pdfborder={0 0 1},backref=false,colorlinks=false]
 {hyperref}
\hypersetup{
 colorlinks,linkcolor=blue,citecolor=blue,urlcolor=blue}
\renewcommand{\vec}[1]{\boldsymbol{#1}}

\begin{document}
\title{Enhancement of Spin-charge Conversion in Dilute Magnetic Alloys\\
by  Kondo Screening
}

\author{Chunli Huang}

\affiliation{Department of Physics, The University of Texas at Austin, Austin,
Texas 78712,USA}

\author{Ilya V.~Tokatly}
\affiliation{Nano-Bio
  Spectroscopy group and European Theoretical Spectroscopy Facility (ETSF), Departamento de Pol\'imeros y Materiales Avanzados: F\'isica, Qu\'imica y Tecnolog\'ia, Universidad del
  Pa\'is Vasco, Av. Tolosa 72, E-20018 San Sebasti\'an, Spain}

\affiliation{IKERBASQUE, Basque Foundation for Science, E-48011 Bilbao, Spain}
\affiliation{Donostia International Physics Center (DIPC), Manuel de Lardizabal 4, E-20018 San Sebastian, Spain}

\author{Miguel A. Cazalilla}
\affiliation{Donostia International Physics Center (DIPC), Manuel de Lardizabal 4, E-20018 San Sebastian, Spain}
\affiliation{Yukawa Institute for Theoretical Physics, Kyoto University, Kyoto 606-8502, Japan}

\begin{abstract}
We derive a kinetic theory capable of
dealing both with large spin-orbit coupling and Kondo screening in dilute 
magnetic alloys. 
We obtain the  collision integral non-perturbatively and uncover a contribution proportional to the momentum derivative of the impurity scattering S-matrix. The latter yields an important correction to the spin diffusion and spin-charge conversion coefficients, and fully captures the so-called side-jump process without resorting to the Born approximation (which fails for resonant scattering), or to otherwise heuristic derivations.
We apply our kinetic theory to a quantum impurity model with strong spin-orbit, which captures the most important features of Kondo-screened Cerium impurities
in alloys such as La$_{1-x}$Cu$_6$. We find 1) a large zero-temperature spin Hall conductivity that depends solely on the Fermi wave number and 2) a transverse spin diffusion mechanism that modifies the standard Fick's diffusion law. 
 Our predictions can be readily verified by standard spin-transport measurements in metal alloys with Kondo impurities.
 \end{abstract}

\maketitle

\textit{Introduction--}
Topological materials with strong electronic correlation and large spin-orbit coupling (SOC) \cite{witczak2014correlated,dzero2010topological,jackeli2009mott,PhysRevLett.100.156401,PhysRevLett.113.106401,dzero2010topological,dzero2012theory,dzero2016topological} are promising platforms for the realization of exotic phases of matter, with potential applications in spintronics \cite{bader2010spintronics,RevModPhys.76.323,ramaswamy2018recent,khang2018conductive,fukami2016magnetization,fan2014magnetization}. 
One recent example that is being intensively researched are Weyl-Kondo semimetals in heavy fermion compounds~\cite{lai2018weyl,dzsaber2017kondo,grefe2020weyl}.
Below the (Kondo) coherence temperature, the local magnetic moments in these materials form a topologically non-trivial band with Weyl points pinned at the Fermi level. The existence of the latter is believed to lead to the giant Hall effect in $\text{Ce}_3\text{Bi}_4\text{Pd}_3$ ~\cite{dzsaber2018giant}. In the opposite limit of a periodic arrangement that yields a coherent band structure, a giant spin Hall effect has  been observed in disordered alloys of FePt/Au \cite{seki2008giant}. A theoretical explanation has been put forward for the latter in terms of an orbital-dependent Kondo effect.~\cite{guo2009enhanced,Shick_PhysRevB.84.113112}

 Driven by these exciting developments,  in this letter we report a kinetic theory capable of describing, the coupled spin and charge transport in dilute magnetic alloys with Kondo screened
 impurities as well as other types of impurities.
 Note that, unlike ordinary potential scattering, Kondo screening is a strong correlation phenomenon that arises from the antiferromagnetic exchange interaction between a local magnetic moment and the conduction electrons.  
 The screening of an impurity magnetic moment results in a strong (often resonant) scattering of the conduction electrons at the Fermi energy when the temperature is lower than the Kondo temperature. Under such conditions and in the presence of large SOC, we have found that the spin-Hall effect is substantially enhanced and the spin diffusion coefficients become spin-anisotropic. The abundance of dilute magnetic alloys allow our predictions to be readily tested by existing experimental techniques (e.g.~\cite{TakMae2008}). 
Below, we develop a model that can be applied to alloys containing rare earth impurities, such as Cerium in Ce$_{x}$ La$_{1-x}$Cu$_6$ for which a robust Kondo effect has been observed in electrical resistivity measurements~\cite{Sumiyama1986}, but no spin transport measurements have been carried out so far to the best of our knowledge.

The most direct  manifestations of SOC in transport experiments are the anomalous Hall effect (AHE) \cite{RevModPhys.82.1539} and the spin Hall effect (SHE) \cite{sinova2015spin}.
Depending on the origin of SOC, one usually distinguishes between intrinsic and extrinsic contributions to the transverse conductivity. The former is related to SOC generated by the periodic crystal potential of the lattice and encoded in the electronic band structure, while the latter originates from the SOC of randomly distributed impurities. In turn, the extrinsic contribution is further divided into two distinct mechanisms: skew-scattering and side-jump. Skew-scattering arises due to the angular asymmetry of the scattering cross section and therefore it can be readily incorporated in the collision integral of the kinetic (Boltzmann-like) equation. Among all mechanisms, the side-jump \cite{PhysRevLett.29.423,PhysRevB.2.4559,NozLew1973,levy1988extraordinary,TseSar2006,HanVig2006,sinitsyn2006coordinate,PhysRevB.81.125332} appears to be the least understood. Physically it can be attributed to a spin-dependent transverse shift (jump) of a wave packet scattered off the impurity. Since this effect does not show up in the scattering cross section, its inclusion in the kinetic theory is by no means straightforward.
It is typically done heuristically by defining a coordinate ``jump'' $\delta \vec{r}$ of a wave packet, introducing the related anomalous velocity and a modified carrier energy dispersion, and incorporating these ingredients into the kinetic equation using reasonable, but still non-rigorous arguments \cite{levy1988extraordinary,Fert_resonant1}
\footnote{See Ref.~\cite{SM} for a detailed discussion of why the anomalous velocity derived in Ref.~\cite{levy1988extraordinary} is only correct in the Born approximation and for disorder potentials without `vertex corrections'.}.
On the other hand, a formal justification of the above procedure and/or derivations of the side-jump contribution from the rigorous quantum kinetic theory practically always rely on the lowest order Born approximation \cite{PhysRevB.81.125332}. 
Such an approach fails for magnetic impurities of heavy elements in the Kondo regime when neither scattering nor SOC can be considered weak. This motivates us to construct a kinetic theory to properly describe all extrinsic mechanisms (including side-jump) self-consistently without resorting to any finite order Born approximation.
We achieved this by computing the  lesser impurity self-energy to first order in spatial derivative (but all orders in disorder potential strength). This gives rise to an additional collision integral $\hat{I}_1$ to the standard kinetic equation. When the kinetic equation is solved in the presence of $\hat{I}_1$, the side-jump correction to spin-Hall conductivity and diffusion constants follows automatically without any heuristic arguments.

Our theory predicts that the standard Fick's law of spin diffusion is modified by SOC when we go beyond the Born approximation: in addition to the standard Laplacian operator $\nabla^2{\bf s}$, the diffusion operator acquires a new term $\sim \nabla (\nabla\cdot{\bf s})$ because SOC breaks the spin-rotation symmetry. This correction occurs at second order in SOC magnetic field.

\textit{Kinetic Theory:---}
We start from the Kadanoff-Baym equation for the nonequilibrium Green functions. Keeping only leading order terms in the impurity density $n_{im}$ we sum up exactly the entire Born series and perform gradient expansion, which allows us to obtain the following kinetic equations \cite{SM} for the spin-density matrix $\hat{n}_{\vec{p}}\equiv \hat{n}_{\vec{p}}(\vec{r},t)$:
 \begin{align}
\partial_{t} \hat{n}_{\vec{p}}  +\vec{v}_{\vec{p}}\cdot\nabla_{\vec{r}}\hat{n}_{\vec{p}} +i [  \Sigma_{\vec{p}}^{H}, \hat{n}_{\vec{p}}] =\hat{I}_0[\hat{n}_{\vec{p}}]+ \hat{I}_{1}[ \hat{n}_{\vec{p}}]. \label{eq:BTE}
\end{align}
Here $\epsilon_p=p^2/(2m^*)$ is the single-particle energy dispersion, $\vec{v}_{\vec{p}}= \nabla_{\vec{p}}\epsilon_p$, and $\Sigma_{\vec{p}}^{H} = n_{im}(T^{R}_{\vec{p}\vec{p}}+T^{A}_{\vec{p}\vec{p}})/2$ is the mean-field generated by impurities, where $T_{\vec{pk}}^{R(A)}$ is the exact single-impurity retarded (advanced) scattering $T$-matrix. The $T$-matrix also determines the collision integrals in the right-hand side of Eq.~\eqref{eq:BTE}, which describes, amongst other effects, the momentum and spin relaxation caused by impurity scattering:
\begin{widetext}
\begin{eqnarray}
&&\hat{I}_0[\hat{n}_{\vec{p}}]_{\alpha,\beta}= 2\pi n_{im}\sum_{\vec{k}}\delta(\epsilon_{\vec{p}}-\epsilon_{\vec{k}})\bigg(T_{\vec{p}\vec{k}}^{R}\,\hat{n}_{\vec{k}}\,T_{\vec{k}\vec{p}}^{A}-\frac{1}{2}\big\{ T_{\vec{p}\vec{k}}^{R}T_{\vec{k}\vec{p}}^{A}, \hat{n}_{\vec{p}}\big\}\bigg)_{\alpha \beta} \longrightarrow -\frac{n_{im}}{2\pi}\sum_{\vec{k}} \Lambda_{\alpha\beta,\gamma\delta}(\vec{p},\vec{k}) \, \delta\hat{n}_{\vec{k}, \gamma\delta} \label{eq:I0},\\
&&\hat{I}_1[n_{\vec{p}}]_{\alpha,\beta}= \pi n_{im} \sum_{\vec{k}}\delta(\epsilon_{\vec{p}}-\epsilon_{\vec{k}})\,i \bigg( T_{\vec{p}\vec{k}}^{R}\,
(\nabla_{\vec{r}}    \hat{n}_{\vec{k}})
\cdot\left(\vec{D}_{\vec{pk}}T_{\vec{kp}}^{A}\right) - \mathrm{h.c.~}\bigg)_{\alpha \beta}
\longrightarrow   \pi n_{im}\sum_{\vec{k}} \vec{V}_{\alpha\beta,\gamma\delta}(\vec{p},\vec{k}) \cdot \nabla_{\vec{r}} \delta\hat{n}_{\vec{k}, \gamma\delta} ,
 \label{eq:I1}
\end{eqnarray}
\end{widetext}
where $\vec{D}_{\vec{pk}}=\nabla_{\vec{p}} + \nabla_{\vec{k}}$ is a momentum shift generator. 

Eqs.~\eqref{eq:I0} and \eqref{eq:I1} are the main results of this work and provide the basis for our combined treatment of strong scattering resulting from Kondo screening and large SOC. Eq.~\eqref{eq:I0} is the matrix generalization \cite{lifshitz2009,chunli2016} of the golden-rule collision integral derived by Luttinger and Kohn \cite{PhysRev.109.1892}, which has a Lindbladian structure often encountered in open quantum systems \cite{breuer2002theory}. As we explain in what follows, the leading gradient correction to the collision integral, $\hat{I}_1[n_{\vec{p}}]$ in Eq.~\eqref{eq:I1}, accounts for the side-jump mechanism. Indeed, 
the role of $\hat{I}_1$ is twofold. First, because $\hat{I}_1\sim \nabla_{\vec{r}} \hat{n}_{\vec{k}}$, it  renormalizes the velocity entering the drift term of Eq.~\eqref{eq:BTE}, thus generating an anomalous contribution to the current as $\vec{D}_{\vec{pk}}T^A_{\vec{pk}}=i \langle \vec{p}| [T^{A},\vec{r}] |\vec{k}\rangle$ which has its origin in the impurity potential.  Second, in the presence of an external field that can be introduced by trading the density for the  electro-chemical potential (\textit{i.e.} $\vec{\nabla}\rho= \sum_{\vec{k}} \mathrm{Tr}\:  \nabla_{\vec{r}} \hat{n}_{\vec{k}} \to  N_F \nabla_{\vec{r}} \mu = e  N_F \vec{E}$, where  $\vec{E}$ is the electric field and $N_F$ is the density of states at the Fermi energy),  it generates a coupling to the electric field, proportional to  $n_{im}$. 
The latter leads to the very special scaling with the impurity concentration of the side-jump contribution to the transport coefficients. In particular, the corresponding contribution to the spin Hall conductivity is independent on $n_{im}$ -- the well known signature of the side-jump mechanism \cite{RevModPhys.82.1539,sinova2015spin}.
 When the $T$-matrix is replaced with the \textit{bare} impurity potential, $\pi n_{im} \vec{V}$ in Eq.~\eqref{eq:I1} becomes the anomalous velocity derived in Ref.~\cite{PhysRevLett.29.423} within the Born approximation.

In the most practically important linear regime, the deviation of $\hat{n}_{\vec{k}}$ from the Fermi function $n_{F}(\epsilon_{\vec{k}})$ is bound to the Fermi surface (FS), $\hat{n}_{\vec{k}}-n_{F}=\delta(\epsilon_{\vec{k}}-\epsilon_F)\delta \hat{n}_{\vec{k}}$, where $\epsilon_F$ is the Fermi energy. In this regime the collision integrals $\hat{I}_0$ and $\hat{I}_1$ simplify as shown by arrows in Eqs.~\eqref{eq:I0} and \eqref{eq:I1}, respectively. The fourth rank tensors $\check{\Lambda}(\vec{p},\vec{k})$ and $\check{\vec{V}}(\vec{p},\vec{k})$ depend only on directions of momenta on the FS and act as super-operators on the FS density matrix $\delta \hat{n}_{\vec{k}}$. They are conveniently expressed in terms of the scattering $S$-matrix $S_{\alpha\beta}(\vec{p},\vec{k})$ and the on-shell $T$-matrix $t_{\alpha\beta}(\vec{p},\vec{k})=\frac{1}{2\pi i}[\delta_{\vec{p}\vec{k}}\delta_{\alpha\beta}-S_{\alpha\beta}(\vec{p},\vec{k})]\equiv\delta(\epsilon_{\vec{p}}-\epsilon_{\vec{k}})T_{\alpha\beta}(\vec{p},\vec{k})\,$:
\begin{equation} \label{eq:I0-S}
\Lambda_{\alpha\beta,\gamma\delta}(\vec{p},\vec{k})= \delta_{\vec{pk}}\delta_{\alpha\gamma}\delta_{\beta\delta} 
-S_{\alpha\gamma}(\vec{p},\vec{k})S_{\beta\delta}^{*}(\vec{p},\vec{k}) , \end{equation}
\begin{equation}
\vec{V}_{\alpha\beta,\gamma\delta}(\vec{p},\vec{k}) = t_{\alpha\gamma}(\vec{p},\vec{k})
i (\overrightarrow{D}_{\vec{p}\vec{k}}- \overleftarrow{D}_{\vec{p}\vec{k}})
t_{\beta\delta}^{*}(\vec{p},\vec{k}). \label{eq:I1-S}
\end{equation}
$\Lambda_{\alpha\beta,\gamma\delta}$ has a typical form of a relaxation super-operator commonly used to describe spin decoherence in atoms and molecules \cite{liu1975theory,dyakonov1979decay}. The vector-valued ``velocity super-operator'' $\vec{V}_{\alpha\beta,\gamma\delta}$ is related to the momentum-gradient
of the scattering phase and thus to the coordinate shift of the scattered wave packet. 
In fact, Eqs.~\eqref{eq:I1-S} and \eqref{eq:I1} provide a precise non-perturbative definition of the side-jump process and clarify the way it enters a consistent quantum kinetic theory.

\textit{Diffusive limit--}
In a typical transport situation the momentum relaxation length (mean free path) is much shorter than characteristic scales of space inhomogeneities. In this so called diffusive regime the distribution function $\delta\hat{n}_{\vec{k}}$ becomes almost isotropic and is fully determined by its 0th $\sum_{\vec{k}}\delta\hat{n}_{\vec{k}}$ and 1st $\sum_{\vec{k}}\vec{k}\delta\hat{n}_{\vec{k}}$ moments:
\begin{equation}
 \label{delta-n}
N_{F}\delta\hat{n}_{\vec{k}}\approx \rho\mathbb{1} + s_a\sigma_a +3k_i(g_{i0}\mathbb{1}+g_{ia}\sigma_a)v_F^{-1},
\end{equation}
where $\sigma_a$ are Pauli matrices, $\mathbb{1}$ is a 2$\times$2 unit matrix, $N_F$ ($v_F$) is the density of states (Fermi velocity) at the FS, $\rho$ and $\vec{s}$ are the charge and spin densities, and $g_{i0}$ and $g_{ia}$ are charge and spin parts of the 1st moment. By substituting Eq.~\eqref{delta-n} into the kinetic equation and taking its 0th and 1st moments we arrive at a system of equations  coupled by the moments of super-operators $\check{\Lambda}$ and $\check{\vec{V}}$. Then, elimination of $g_{i0}$ and $g_{ia}$ yields a closed set of equations of motion for $\rho$ and $\vec{s}$ -- the charge-spin diffusion equations which we now derive explicitly.

To be specific, we assume \textit{isotropic} disorder potential which leads to a $T$-matrix that is invariant under time-reversal, parity and the full spin-orbit rotations~\cite{taylor2006scattering}. With
these assumptions, we diagonalized the kinetic equation by taking suitable linear-combinations of the \textit{ansatz} (r.h.s of Eq.~\eqref{delta-n}) and solution can be obtained without assuming the collision integral is small
 (see Ref.~\cite{SM} for full details). 
The 0th moment of the kinetic equation yields the charge and spin continuity equations,
\begin{equation}
 \label{continuity}
 \partial_{t}\rho+\partial_{j}\mathbb{J}_{j}=0, \qquad 
 \partial_{t}s_{b}+\partial_{j}\mathbb{J}_{jb}=- s_{b}/\tau_{s},
\end{equation}
where the charge $\mathbb{J}_{j}$ and spin $\mathbb{J}_{jb}$ currents are linear combinations of the charge and spin 1st moments of $\delta\hat{n}_{\vec{k}}$~\cite{SM}.  The spin relaxation time $\tau_s$ in Eq.~\eqref{continuity} is determined by the angular average of the relaxation super-operator $\check{\Lambda}$, $\tau_s^{-1}\sim n_{im}{\rm tr}\langle\sigma_a\check{\Lambda}\sigma_a\rangle$.

 By taking the 1st moment of the kinetic equation, and solving it for the 1st moments of $\delta\hat{n}_{\vec{k}}$, we relate the currents to charge and spin density gradients~\cite{SM}:
\begin{align}
\mathbb{J}_{j} & =-D_{c}\partial_{j}\rho-D\theta_{sH}\,\epsilon_{jka}\,\partial_k s_a \label{eq:Jc-final}\\
\mathbb{J}_{jb}  & =-\sum_{m=0}^2D_{m} P_{jb}^{m}-D\theta_{sH} \, \epsilon_{jkb} \,\partial_{k}\rho, \label{eq:Js-final}
\end{align}
where $P_{jb}^{m}$ are irreducible tensors of spin gradients: 
\begin{align} \nonumber
P_{ja}^{m=0}=\frac{1}{3}\delta_{aj}\partial_{i}s_{i} 
\;\;  , \; \;
P_{ja}^{m=1}=\frac{1}{2}(\partial_{j}s_{a}-\partial_{a}s_{j}) , \nonumber\\
P_{ja}^{m=2}=\frac{1}{2}(\partial_{j}s_{a}+\partial_{a}s_{j})- \frac{1}{3}\delta_{aj}\partial_{i}s_{i} \quad 
\label{eq:P}
\end{align}
The diffusion currents are parameterized by the spin Hall angle $\theta_{sH}$ 
, the charge diffusion constant $D_c$ and three spin diffusion constants $D_{m}$, which are related to different angular averages of the super-operators $\check{\Lambda}$ and $\check{\vec{V}}$ \cite{SM},
\begin{align}
\theta_{sH} & =\frac{(1-\Omega_{c}-\Omega_{1})\theta_{sk}-\Omega_{cs}-\Omega_{sc}\gamma_{1}}{ \gamma_{1}+2\theta_{sk}^{2}}
\label{eq:hall_con}\\
D_{c} & =D\frac{\gamma_{1}(1-2\Omega_{c})+ 4\theta_{sk}\Omega_{cs}}{ \gamma_{1}+2\theta_{sk}^{2}}\\
D_{1} & =D\frac{(1-2\Omega_{1})+4\theta_{sk}\Omega_{sc}}
{ \gamma_{1}+2\theta_{sk}^{2}} \label{eq:D1} \\
D_{m} & =D({1-2\Omega_{m}})/{ \gamma_{m}} \label{eq:D0&D2}\; , \qquad m=0,2
\end{align}
Here the coefficients $\Omega_c$, $\Omega_{m}$, $\Omega_{cs}$ and $\Omega_{sc}$ are generated by the velocity super-operator $\check{\vec{V}}$, e.g.~$\Omega_{sc}\sim n_{im}{\rm tr}\langle\bm{\sigma}\cdot(\vec{k}\times\check{\vec{V}})\mathbb{1}\rangle$. Physically, $\Omega_c$ and $\Omega_{m}$ renormalize the effective charge and spin velocities, while $\Omega_{cs}$ and  $\Omega_{sc}$  account for the side-jump mechanism of the charge-to-spin conversion. Finally, $D=\frac{1}{3}v_F^2\tau_{\rm tr}$, $\gamma_m$, and $\theta_{sk}$, together with $\tau_s$ in Eq.~\eqref{continuity} parameterize the super-operator $\check{\Lambda}$. The explicit formula for all these coefficients are provided in~\cite{SM}.

The above expressions provide the complete solution to kinetic theory in the diffusive limit.
Equations \eqref{continuity}, \eqref{eq:Jc-final}, and \eqref{eq:Js-final} describe the diffusion of  spin and charge for any value of single-impurity potential strength in the dilute limit.
The linear response to an external field can be read off from the diffusion equations using the Einstein relation, \textit{i.e.} by introducing an electric field as described under Eq.~\eqref{eq:I1}, both the charge and the transverse spin Hall conductivity can be obtained from Eqs.~\eqref{eq:Jc-final} and \eqref{eq:Js-final}, which yields $\sigma_{c}=e^2 D_c N_F $ and  $\sigma_{sH}=e D \theta_{sH} N_F$. 

Instead of writing the spin current in terms of the coefficients $D_m$, it is also instructive to separate explicitly its divergence-less part of $\mathbb{J}_{jb}$ and rewrite Eq.~\eqref{eq:Js-final} as follow:
\begin{eqnarray}
 \nonumber
  \mathbb{J}_{jb}&=& -D_s^T\partial_js_b - (D_s^L-D_s^T)\partial_bs_j \\
  &-& \kappa (\partial_bs_j - \delta_{jb}\partial_ks_k) - D\theta_{sH}\,\epsilon_{jkb}\partial_{k}\rho,
  \label{L-T-kappa}
\end{eqnarray}
where $D_{s}^T=(D_{1}+D_{2})/2 $, $D_{s}^L= (D_{0}+2D_{2})/3$, and $\kappa=(D_2-D_0)/3$. The third term entering this equation with the coefficient $\kappa$ is the ``swapping current'' predicted in \cite{lifshitz2009}. Since the swapping current and the spin Hall current have zero divergence, only the first line in  Eq.~\eqref{L-T-kappa} contributes to the bulk spin diffusion equation,
\begin{equation} \label{eq:bloch}
\partial_t \vec{s} - D_{s}^T \nabla^2 \vec{s} - (D_{s}^L-D_{s}^T) \nabla(\nabla \cdot \vec{s})= -\vec{s}/\tau_s.
\end{equation}
Besides the usual Fick's term $\sim\nabla^2\vec{s}$ \cite{abanin2009nonlocal,chunli2017}, the diffusion operator above contains an additional term $\sim\nabla(\nabla\cdot\vec{s})$ that breaks the spin-rotation symmetry while preserving the full space+spin rotation invariance respected by SOC. Physically, the new term leads to different diffusion laws for the transverse $\vec{s}^T$ (with $\nabla \cdot \vec{s}^T=0$) and longitudinal $\vec{s}^L$ (with $\nabla \times \vec{s}^L=0$) components of the spin density. In fact, $D_s^T$ and $D_s^L$ are the diffusion constant for $\vec{s}^T$ and $\vec{s}^L$, respectively. To the leading order in SOC, we find $D_s^L\approx D_s^T$, so a sufficiently large SOC is needed to make the effect observable as we discuss next.
\begin{figure}[t]
\includegraphics[width=0.85\columnwidth]{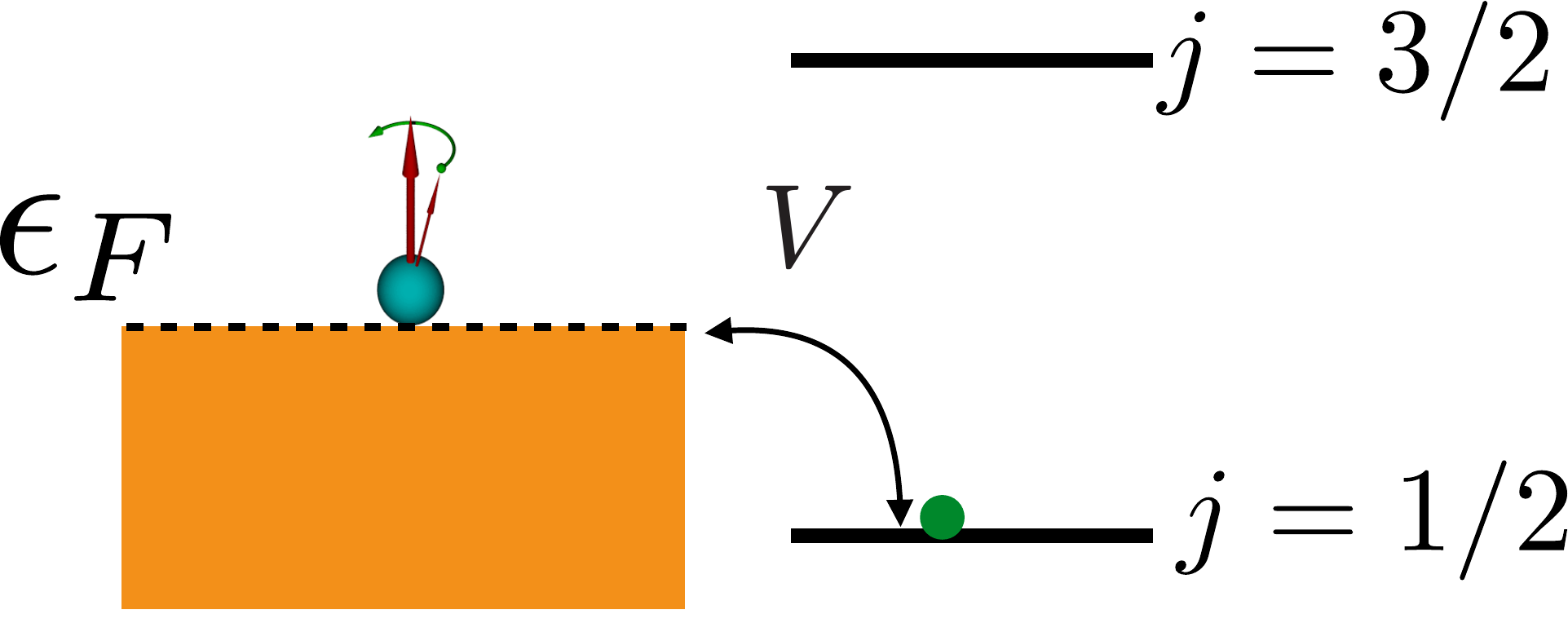}
\caption{Sketch of the  minimal quantum impurity model to which we have applied our kinetic theory. The impurity contains a single electron in an $l=1$ orbital that, by virtue of strong spin-orbit coupling,  splits into a $j=1/2$ doublet and $j=3/2$ a quartet. Strong electron correlation leads to the formation of a local moment. Kondo screening of the latter by  the conduction electrons induces a scattering phase-shift  $\eta_{1}=\pi/2$ at the Fermi energy.
See~\cite{SM} for a detailed explanation of how this model captures some essential features of Ce impurities in alloys like Ce$_x$La$_{1-x}$Cu$_6$ for $x < 0.7$.} 
\label{fig:schematic}
\end{figure}

%

\textit{Quantum impurity model--}
We now use a simple quantum impurity model \cite{SM}
to demonstrate the effect of Kondo screening on spin Hall conductivity $\sigma_{sH}$ and the anisotropic spin diffusion parameter $D_s^L/D_s^T$.
 This model is intended to capture some of the basic features of the Ce impurities in the Kondo-screened regime in dilute alloys such as Ce$_x$La$_{1-x}$Cu$_6$ with $x < 0.7$ \cite{HewsonKondo,Kawakami1986}.
Since Cu has negligible SOC, we can use Eq.~\eqref{eq:BTE} to describe (extrinsic) spin-transport in this alloy.
The ground state of a single f-electron in the Ce atom is a doublet which is separated by $\sim 100 $ K from a quartet~\cite{HewsonKondo,Kawakami1986} due to the crystal environment.
We model this low-lying multiplet structure using a $l=1$ orbital that is split by an effective SOC into a doublet with $j=1/2$ and a quartet with $j=3/2$, as shown in Fig.~1. Thus, we find that, contrary to conventional wisdom \cite{RevModPhys.82.1539}, the spin Hall conductivity arises entirely from the side-jump mechanism when the T-matrix is dominated by a single non s-wave scattering channel.

In the Kondo-screened regime (\textit{i.e.}~$T \ll T_K \sim 1$ K), the on-shell $T$-matrix at the FS of this quantum impurity model can be derived using standard many-body technique described in Ref.~\cite{SM} and the result is
\begin{align}   \label{eq:tmatrix_maintext}
\hat{T}^{R}_{\vec{k}\vec{p}} =
&-\frac{
  e^{i\eta_{0}} \sin \eta_0 +
(e^{i\eta_{1}} \sin \eta_1 + 2e^{i\eta_{2}} \sin \eta_2)\hat{\vec{p}}\cdot \hat{\vec{k}}
 }{\pi N_F}\mathbb{1}   \nonumber \\
&- \left[ \frac{ e^{i\eta_{1}} \sin \eta_1-e^{i\eta_{2}} \sin \eta_2 }{\pi N_F} \right] i (\hat{\vec{k}} \times \hat{\vec{p}})\cdot \vec{\sigma}.
\end{align}
Here $\eta_{1}$ ($\eta_{2}$) is the scattering phase-shifts of the $l=1, j=1/2$ and ($l=1, j=3/2$) channel shown in Fig.~1; $\eta_0$ is the phase-shift of the usual $s$-wave $l=0,j=1/2$ channel.  The doublet ground state of the Ce is Kondo-screened and therefore Ce behave as non-magnetic scatterer that induces a resonant scattering phase-shift. Thus, in our model conduction electrons undergo the strongest scattering in the $\eta_1$-channel for which $\eta_1 = \pi/2$. If we set $\eta_0=\eta_2=0$, then $\theta_{sH}=-\Omega_{cs}-\Omega_{sc}$, which yields a spin Hall conductivity that arises entirely from the side-jump mechanism:
\begin{equation} \label{eq:sigma_xy}
\sigma_{sH}  = -\frac{e D (\Omega_{cs}+\Omega_{sc}) N_F}{\hbar}  = \frac{4 }{9\pi^2 } \frac{e k_F}{\hbar}.
\end{equation}
We have reintroduced $\hbar$ above. It is interesting to point out in the single scattering channel limit, $\sigma_{sH}$  does not depend on the impurity density and the specific value of $\eta_1$. This is because the $n_{im}$ and $\eta_1$ dependence of $D=\frac{1}{3}v_F^2\tau_{\text{tr}}\propto (n_{im}\sin^2\eta_1)^{-1}$  exactly canceled by $\theta_{sH} \propto \Omega_{cs} \propto n_{im}\sin^2 \eta_1$.  Importantly, unlike the case of ordinary impurities with $d$  orbitals~\cite{Fert_resonant1,Fert_resonant2,Fert_resonant3} where the relationship between $\eta_1$ and $\eta_2$ is determined by SOC, in our model $\eta_1=\pi/2$ is determined by the  Kondo screening. Without the Kondo-screening  $\sigma_{sH}$ becomes a complicated function of the three phase shifts $\eta_0, \eta_1, \eta_2$ which we now study.
When the two additional channels $\eta_0$ and $\eta_2$
are weakly coupled and therefore  small in magnitude. We find Eq.~\eqref{eq:sigma_xy} receives a skew-scattering contribution:
\begin{equation} \label{eq:skew}
\sigma^{sk}_{sH} \simeq -\frac{\eta_0 }{12\pi}  \left(\frac{n_c}{n_{im}} \right) \left( \frac{e k_F}{\hbar} \right),
\end{equation}
where  $n_c=k_F^3/3\pi^2$ is the carrier density. The ratio of Eq.~\eqref{eq:sigma_xy} to Eq.~\eqref{eq:skew} is $ \approx 2\eta_0^{-1}(n_{im}/n_c)$.  Numerically,  $e \sigma^{sJ}_{sH}\simeq 2.72 \times 10^{-6} k_F \mathrm{Ohm}^{-1}$. If we use the standard estimate for $\eta_0\simeq 0.1$~\cite{Fert_resonant1,guo2009enhanced,Shick_PhysRevB.84.113112},
then Eq.~\eqref{eq:sigma_xy} becomes comparable in magnitude to $\sigma^{sk}_{sH}$
for $n_{im}/n_c\simeq 5\%$,  for which the resistivity still shows the low temperature saturation characteristic of isolated Kondo-screened impurities.~\cite{Sumiyama1986}.

Finally, let us compute the correction to the na\"ive Fick's law by calculating the deviation of $D_s^L/D_s^T$ from unity.
In the limit where the doublet is Kondo screened $\eta_1=\pi/2$, and the other two orbitals are weakly coupled (i.e. $|\eta_0|,|\eta_2|\ll 1$), we obtain
\begin{equation} \label{eq:Ds_ratio}
    \frac{D_s^L}{D_s^T} \simeq  1 - \frac{8 n_i}{3\pi n_c} \left(\frac{\eta_0}{3}+ \eta_2 \right) + \frac{4 \eta^2_0}{9}.
\end{equation}
It is interesting to point out that $\frac{D_s^L}{D_s^T}-1 \propto B^2$ where $B$ is the spin-orbit magnetic field defined by the square bracket in Eq.~\eqref{eq:tmatrix_maintext}.
When we assume all the phase shifts are small, i.e. $|\eta_{0,1,2}|\ll 1$, the leading corrections to $D_s^L/D_s^T$ are third order in the phase shifts so the spin-anisotropy cannot be captured by the first Born approximation.

Equations  \eqref{eq:bloch}, \eqref{eq:sigma_xy} and \eqref{eq:Ds_ratio} demonstrate that spin-charge conversion mechanisms can be both quantitatively and qualitatively modified as a consequence of the strong scattering induced in one of the scattering channels by Kondo screening 
\footnote{At finite but small temperature $T \lesssim T_K$, inelastic scattering related to the polarization of Kondo screening cloud \cite{nozieres1974fermi,HewsonKondo,SM}  leads corrections of order $(T/T_K)^2$ to the kinetic coefficients which will be studied elsewhere~\cite{in-prep}.}.

\textit{Summary and Discussion--}
We have developed a kinetic theory that provides a general framework to study spin transport in alloys containing dilute random ensembles of impurities with $d$ and $f$ orbitals.
Scattering with such impurities is treated non-perturbatively, allowing us to deal with the strong scattering of conduction electrons on the Fermi surface coupled with strong local spin-orbit (SOC). We have reported an analytical  solution of the kinetic equations for a rotationally invariant system and applied it to simple quantum impurity model designed to capture the essential features of (Kondo-screened) Cerium impurities in alloys such as Ce$_{x}$
La$_{1-x}$Cu$_6$ with $x< 0.7$. 
We find the combination of strong scattering and local SOC lead to a large contribution to the spin Hall conductivity  $\sigma_{sH}$ that stems entirely from the  side-jump and in the limit where interference with other channels can be neglected takes a value that depends only on the Fermi wave number. In addition, our non-perturbative treatment of impurity scattering allows us to show that the spin diffusion coefficients is spin-anisotropic.

The above predictions can be readily tested in spin-valve devices where the spin-current is injected from a ferromagnetic contact along different directions, thus allowing to measure the different spin diffusion lengths associated with to $D^{L,T}_s$ as well as the spin Hall conductivity $\sigma_{sH}$.  When the injected spin is polarized in the direction parallel (perpendicular) to the direction of the current, it measures the longitudinal $D_{s}^L$ (transverse $D_{s}^T$) spin diffusion constant. Due to SOC, $D_{s}^L\neq D_{s}^L$ and this will be the most direct test of our theoretical predictions.

\textit{Acknowledgements--}
C.H.~acknowledges useful discussion with Nemin Wei and Qian Niu. C.H. thanks Donostia International Physics Center for hospitality and acknowledges support from Spanish Ministerio de Ciencia, Innovaci\'{o}n y Universidades (MICINN) (Project No.~FIS2017-82804-P). I.V.T. acknowledges support by Grupos Consolidados UPV/EHU del Gobierno Vasco (Grant No. IT1249-19). MAC acknowledges the support  of Ikerbasque (Basque Foundation for Science). The work of MAC was also carried out by joint research in the International Research Unit of Quantum Information, Kyoto University.

\bibliographystyle{ieeetr}
\bibliography{ref.bib}

\clearpage
\pagebreak
\newpage

\widetext
\begin{center}
\textbf{\large Supplemental Materials: Enhancement of Spin-charge Conversion in Dilute Magnetic Alloys by Kondo Screening}
\end{center}
\setcounter{equation}{0}
\setcounter{figure}{0}
\setcounter{table}{0}
\setcounter{page}{1}
\makeatletter
\renewcommand{\theequation}{S\arabic{equation}}
\renewcommand{\thefigure}{S\arabic{figure}}
\renewcommand{\bibnumfmt}[1]{[S#1]}
\renewcommand{\citenumfont}[1]{S#1}

\vspace{1cm}

This supplementary material is organized as follows. In Section I, we provide the details of the derivation of the kinetic
equation for a ($2 \times 2$) spin-density matrix distribution function in the presence of dilute ensemble of impurities and discuss the form of the collision integral. In Section II, we use the kinetic equation to discuss the correct form side-jump anomalous velocity and compared to previously derived expressions.
In Section III, we provide additional details of the solution of the kinetic equation sketched in the main text for an isotropic impurity potential. We consider the solution in the diffusive limit without assuming the smallness of the impurity spin-orbit coupling (SOC) and potential strength
The solution is characterized by $12$ kinetic coefficients and 
the expressions relating these coefficients to the single-impurity scattering matrix of a simple impurity model are provided in Section III. In section IV, we introduce an isotropic quantum impurity model intended to model rare-earth Cerium impurities in a metal host such as Cooper. We  discuss the low temperature properties
of this quantum impurity model and carefully use symmetry principles to construct the impurity scattering matrix. Except for the description of the model and the last subsection of Sect. IV, a large of the material presented in this section is textbook level and can be skipped by the more specialized readers. It is included here for the sake of pedagogy and completeness.  Finally, in section V, we provide a short proof to show the eigenvalues of the relaxation super-operator are real and non-negative.

\section{derivation of kinetic equation}\label{sec:derivation}

In this section, we derive the kinetic equation given in the main-text
using the non-equilibrium Green's function formalism. The derivation
of kinetic equation from this formalism is a standard tool that has
been widely used in many areas in physics so we shall be brief.

A kinetic theory aims to describe the dynamics of many-particle system
with a distribution function $n_{\vec{p}}(\vec{r},t)$. The time evolution
of distribution function can be written as ($\partial_{t}+v_{\vec{p}}\cdot\nabla_{\vec{r}})n_{\vec{p}}(\vec{r},t)=\mathcal{C}[n_{\vec{p}}(\vec{r},t)]$
where $\text{\ensuremath{\mathcal{C}}}$ is a complicated differential-integral
operator and it inevitably has to be approximated by some small parameters
in a theory. For impure metal in the diffusive limit, $\mathcal{C}$
is usually expanded to zeroth order in spatial non-locality ($\nabla_{r}$) and linear order in 
and $n_{im}$ where $n_{im}$ is the impurity density. In this work,
we go beyond the usual expansion scheme to include term at order $n_{im}\nabla_{r}$.
This term has important consequences for metals with SOC disorder as it describes the  side-jump mechanism.

The first step in formulating a kinetic theory is to use the equation
of motion for a contour-ordered Green's function $\check{G}\left(1,2\right)=-i\langle T_{c}\psi(1)\psi^{\dagger}(2)\rangle$
where $1=\vec{r}_{1},t_{1},\sigma_{1}$ labels the space, time and
spin of a quasi-particle. In this section, we denote all contour-ordered
quantities with a check, e.g. $\check{\Sigma}$. The equations of
motion are the non-equilibrium generalization of the Dyson equation:
\begin{align}
\partial_{t_{1}}\check{G}(1,2)-i\xi_{1}\check{G}(1,2) & =-i\sum_{\sigma=\pm1}\int_{\Omega}d^{3}r_{3}\int_{\gamma_{c}}dt_{3}\;\check{\Sigma}(1,\vec{r}_{3}t_{3}\sigma_{3})\check{G}(\vec{r}_{3}t_{3}\sigma_{3},2)\label{eq:left}\\
\partial_{t_{2}}\check{G}(1,2)-i\check{G}(1,2)\xi_{2} & =-i\sum_{\sigma=\pm1}\int_{\Omega}d^{3}r_{3}\int_{\gamma_{c}}dt_{3}\;\check{G}(1,\vec{r}_{3}t_{3}\sigma_{3})\check{\Sigma}(\vec{r}_{3}t_{3}\sigma_{3},2)\label{eq:right}
\end{align}
where $\xi_{1}=-\nabla_{r_{1}}^{2}/2m^{*}$ is the kinetic energy
and $\check{\Sigma}$ is the self-energy defined on the complex time
contour. Note $\text{\ensuremath{\check{\Sigma}}}=\check{\Sigma}[\check{G}]$
is a functional of $\hat{G}.$ In the above, $\Omega$ is the integration
volume and $\gamma_{c}$ is an integration contour for the complex
time which we take it to be the standard Keldysh-contour. $\check{G}$
contains information about the dynamics of quasi-particle energy spectrum
and the distribution function of quasi-particles. For example, if we subtract Eq.\eqref{eq:left} from \eqref{eq:right} and take the
lesser component on the time-contour, we arrived at equation of motion
for a matrix-valued distribution function $G^{<}\left(1,2\right)=-i\hat{\rho}(1,2)=-i\langle\psi^{\dagger}(2)\psi(1)\rangle$:
\begin{equation}
\left(\partial_{t_{1}}+\partial_{t_{2}}-i(\xi_{1}-\xi_{2})\right)G^{<}=-i\left(\Sigma^{R}\otimes G^{<}+\Sigma^{<}\otimes G^{A}-G^{R}\otimes\Sigma^{<}-G^{<}\otimes\Sigma^{A}\right)\label{eq:Glesser}
\end{equation}
 Above, we have used the convolution notation
$A\otimes B=\sum_{\sigma}\int_{\Omega}dr_{3}\int_{-\infty}^{\infty}dt_{3}A(1,3)B(3,2)$ 
as well as Langreth's rule $(A\times B)^{<} = A^R\otimes B^{<} + A^{<}\otimes B^A$. 
Note the time-integration runs from
$-\infty$ to $\infty$. In the above, $\Sigma^{R(A)}$ and $\Sigma^{<}$
are the retarded (advanced) and lesser self-energy. In order to extract
information about the quasi-particle spectral weight $A=\frac{1}{2i}(G^{>}-G^{<})=\frac{1}{2i}(G^{R}-G^{A})$,
we write down the equation of motion for $G^{>}$ and subtract it
from Eq.\eqref{eq:Glesser}. The result reads:
\begin{equation}
\left(\partial_{t_{1}}+\partial_{t_{2}}-i(\xi_{1}-\xi_{2})\right)A=-\frac{i}{2}\left[\Sigma^{R}+\Sigma^{A}\overset{\otimes}{,}A\right]-\frac{1}{4}\left[\left(\Sigma^{R}-\Sigma^{A}\right)\overset{\otimes}{,}\left(G^{R}+G^{A}\right)\right]=0+O(n_{im})\label{eq:spectrum}
\end{equation}
Eq.~\eqref{eq:Glesser} and \eqref{eq:spectrum} are two coupled
equations of motion. For impurity induced self-energy, we are mainly
interested in describing the effects of collision of quasi-particles 
with impurities on the distribution function  and not on 
the quasi-particle spectrum. 
To the leading (i.e. zeorth) order in impurity density, one
can set the right hand side of Eq.\eqref{eq:spectrum} to zero then
$G^{R(A)}(\omega,\vec{p})=(\omega-\epsilon_{\vec{p}}\pm i\delta)^{-1}$
and $A(\vec{r},t,\vec{p},\omega)=-\pi\delta(\omega-\epsilon_{\vec{p}})$
where $\epsilon_{\vec{p}}=\vec{p}^{2}/2m^{*}$ is the bare quasiparticle
energy.

Next, we substitute $G^{R(A)}(\omega,\vec{p})=(\omega-\epsilon_{\vec{p}}\pm i\delta)^{-1}$
into Eq.~\eqref{eq:Glesser} and express it in  Wigner coordinates:
\begin{equation}
(\partial_{t}+\vec{v}_{\vec{p}}\cdot\partial_{r})G^{<}(\vec{r},t,\vec{p},\omega)+\frac{i}{2}\left[\Sigma^{R}(\vec{p},\omega)+\Sigma^{A}(\vec{p},\omega),G^{<}(\vec{r},t,\vec{p},\omega)\right]=I^{<}(\vec{r},t,\vec{p},\omega)
\end{equation}
\begin{equation}
I^{<}(\vec{r},t,\vec{p},\omega)=2\pi\delta(\omega-\epsilon_{p})\Sigma^{<}(\vec{r},t,\vec{p},\omega)+\frac{1}{2i}\{\Sigma^{R}(\vec{p},\omega)-\Sigma^{A}(\vec{p},\omega),G^{<}(\vec{r},t,\vec{p},\omega)\}
\end{equation}
Note  that, in the above expressions, we have used $\Sigma^{R}=\Sigma^{R}[G^{R}]=\Sigma^{R}(\vec{p},\omega)$, which 
is independent of $\vec{r}$ and $t$, and similarly for $\Sigma^{A}$.
This is in general not true in a superconductor where the spectral
weight is space-time dependent and extra care must be taken in retaining
higher corrections in $I^{<}$. To leading order in impurity density, the lesser Green's function takes the standard form
\begin{equation} \label{eq:lesser_GF}
G^{<}(\omega,\vec{p},\vec{r},t)=-2iA(\vec{r},t,\vec{p},\omega)\hat{n}_{\vec{p}}(\vec{r},t)=2\pi i\delta(\omega-\epsilon_{p})\hat{n}_{\vec{p}}(\vec{r},t),
\end{equation}
where the spectral function $A$ is independent of impurity density.
Next, we perform the $\omega$ integration and arrived at the equation
of motion for the distribution function:
\begin{equation}
(\partial_{t}+\vec{v}_{\vec{p}}\cdot\partial_{r})\hat{n}_{\vec{p}}(\vec{r},t)+\frac{i}{2}\left[\Sigma^{R}(\vec{p},\epsilon_{p})+\Sigma^{A}(\vec{p},\epsilon_{p}),\hat{n}_{\vec{p}}(\vec{r},t)\right]=I^{<}(\vec{r},t,\vec{p},\epsilon_{\vec{p}})
\end{equation}
\begin{equation}
I^{<}(\vec{r},t,\vec{p},\epsilon_{\vec{p}})=-i\Sigma^{<}(\vec{r},t,\vec{p},\epsilon_{\vec{p}})+\frac{1}{2i}\{\Sigma^{R}(\vec{p},\epsilon_{\vec{p}})-\Sigma^{A}(\vec{p},\epsilon_{\vec{p}}),\hat{n}_{\vec{p}}(\vec{r},t)\}\label{eq:coll-int-der}
\end{equation}
The discussion so far follows the standard procedure but  some extra care is needed to keep track of the proper arrangement of the Green's function
and the self-energy matrices. In the following, we evaluate the non-equilibrium
self-energy by impurity density expansion and retain important finite
$\nabla_{r}$ (or $\vec{q}$) correction. The self-energy is most
conveniently computed in plane-wave basis. To leading order in impurity
density, electrons scatter with the same impurity located at $\vec{r}_{a}$
multiple times. The resulting Born series is given by the following expression:

\begin{figure} \includegraphics[width=1\columnwidth]{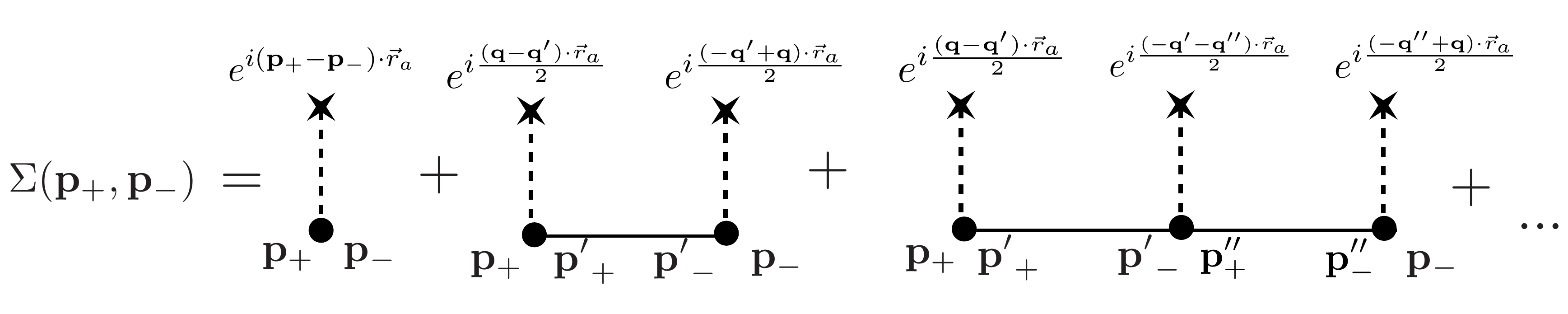}
\caption{Expansion of non-equilibrium self-energy to linear order in density of impurity. Since the (self-consistent) Green function is not necessarily diagonal in momentum, $\langle \psi_{p+q/2}\psi^{\dagger}_{p-q/2}\rangle\neq\langle \psi_{p}\psi^{\dagger}_{p}\rangle \delta_{q,0}$, the wavefunction leaving a vertex (e.g.~$e^{-i(p'+q'/2)\dot r_a})$) does not cancel with the wavefunction entering at the next vertex ($e^{i(p'-q'/2)\dot r_a}$) and this results in addition phase-factors at each vertex. When this series is summed, we obtained a self-energy that is accurate for arbitrary $\vec{q}$.
Here $\vec{p}_{\pm}=\vec{p}\pm\vec{q}/2$, $
{\vec{p}\,}'_{\pm}={\vec{p}\,}' \pm {\vec{q}\,}'/2$
and ${\vec{p}\,}''_{\pm}={\vec{p}\,}''\pm {\vec{q}\,}''/2$.
}
\label{fig:self-energy}
\end{figure}

\begin{align}
\check{\Sigma}(\vec{p}_{+},\vec{p}_{-},\omega,t) & =\frac{1}{\Omega}\sum_{\vec{r}_{a}=1}^{N_{im}}e^{i\vec{q}\cdot\vec{r}_{a}}\left( V_{\vec{p}_{+}\vec{p}_{-}}+\sum_{\vec{p}_{+}'\vec{p}_{-}'}V_{\vec{p}_{+}\vec{p}'_{+}}
\, \check{G}(\vec{p}_{+}',\vec{p}'_{-},,\omega,t;\vec{r}_{a})
\, \check{T}(\vec{p}_{-}',\vec{p}{}_{-},,\omega,t;\vec{r}_{a})\right) \\
  &\equiv \frac{1}{\Omega}\sum_{\vec{r}_{a}=1}^{N_{im}}e^{i\vec{q}\cdot\vec{r}_{a}}\check{T}(\vec{p}_{+},\vec{p}_{-},\omega,t;\vec{r}_{a} )\label{eq:sigma_lesser}
\end{align}
where the second line defines $\check{T}(\vec{p}_{-}',\vec{p}_{-},\omega,t;\vec{r}_{a})$ and
\begin{equation} \label{eq:pm_q_convention}
    \vec{p}_{\pm}=\vec{p}\pm\vec{q}/2 \; ,\; \vec{p}'_{\pm}=\vec{p}'\pm\vec{q}'/2 \; ,\;
    \check{G}(\vec{p}_{+}',\vec{p}'_{-},\omega,t;\vec{r}_{a})\equiv e^{-i\vec{q}'\cdot\vec{r}_{a}}\check{G}(\vec{p}_{+}',\vec{p}'_{-},\omega,t)
\end{equation}
Note that for magnetic impurities 
considered in the main text, we work 
at temperatures much lower than the
Kondo temperature where the magnetic moment has been screened by the conduction
electrons. In the ground state,  the scattering matrix of the impurity corresponds to that of a resonant non-magnetic scatterer, which
nonetheless still has a non-trivial 
at the Fermi energy. Note that in our treatment, this explicit form of this potential is of no importance as it
is replaced by the scattering T-matrix,
which be obtained by other means (see Sec.~\ref{subsec:tmat}). In addition, we have neglected the local interaction that is an  subleading (irrelevant, in the renormalization-group sense) correction to low-energy (renormalization group fixed-point) Hamiltonian that describes the  quantum impurity in the Kondo regime (see discussion in Sec.~\ref{sec:QIM} and Refs.~\cite{Wilson_RevModPhys.47.773,HewsonKondo}).

Since $\check{G}$ is not diagonal in momentum, the electron wave function (plane-wave) leaving a vertex do not cancel with the wave function that enters the next vertex of the diagram, as shown in Fig.\ref{fig:self-energy}. This enforces the spatial coordinate of the Wigner-transformed $\check{G}$ to be located at the impurity position $r_a$.
When the Green's function is diagonal in momentum, such as the
retarded and advanced Green's function,
$G^{R(A)}(\omega,\vec{p})=(\omega-\epsilon_{\vec{p}}\pm i\delta)^{-1}$,
the Born series can be easily summed and the corresponding retarded
(advanced) self-energy reads:
\begin{equation}
\Sigma^{R(A)}(\vec{p}_{+},\vec{p}_{-},\omega,t)=\Sigma^{R(A)}(\vec{p}_{+},\vec{p}_{-},\omega)=\frac{1}{\Omega}\sum_{\vec{r}_{a}=1}^{N_{im}}e^{i\vec{q}\cdot\vec{r}_{a}}T^{R(A)}(\vec{p}_{+},\vec{p}_{-},\omega)\equiv n_{im}T_{\vec{p},\vec{p}}^{R(A)}(\omega)\delta_{\vec{q},0}
\end{equation}
where $T_{\vec{p}\vec{p}}^{R/A}$ is the exact T-matrix generated
by a single impurity that is independent of $\vec{r}_{a}$. It can
be computed, for example, by solving the Lippmann-Schwinger equation
$T^{R}=(1-VG^{R})^{-1}V$. In particular, it obeys the optical
theorem:
\begin{equation}
\frac{\Sigma^{R}(\vec{p},\vec{p},\omega)-\Sigma^{A}(\vec{p},\vec{p},\omega)}{2i}=-\pi n_{im}\sum_{\vec{k}}\delta(\omega-\epsilon_{k})T_{\vec{p}\vec{k}}^{R}(\omega)\,T_{\vec{k}\vec{p}}^{A}(\omega)\label{eq:Sigmar}
\end{equation}
Unlike $\Sigma^{R(A)}$, the lesser self-energy $\Sigma^{<}$ is not
diagonal in $\vec{p}$. If we take the lesser component of Eq.~\eqref{eq:sigma_lesser},
we find that the lesser self-energy has the following important structure
\begin{align} \label{eq:Sigma_less}
\Sigma^{<}(\vec{p}_{+},\vec{p}_{-},\omega,t)&=\frac{1}{\Omega}\sum_{\vec{r}_{a}=1}^{N_{im}}e^{i\vec{q}\cdot\vec{r}_{a}}\sum_{\vec{p}_{+}'\vec{p}_{-}'}T^{R}(\vec{p}_{+},\vec{p}_{+}^{'},\omega)G^{<}(\vec{p}_{+}'\vec{p}'_{-},\omega,t;\vec{r}_{a})T^{A}(\vec{p}_{-}',\vec{p}_{-},\omega) \\
&= \frac{1}{\Omega}\sum_{\vec{r}_{a}=1}^{N_{im}}\sum_{\vec{p}_{+}'\vec{p}_{-}'}
e^{i(\vec{q}-\vec{q'})\cdot\vec{r}_{a}}
T^{R}(\vec{p}_{+},\vec{p}{}_{+}^{'},\omega)G^{<}(\vec{p}_{+}'\vec{p}'_{-},\omega,t)T^{A}(\vec{p}_{-}',\vec{p}{}_{-},\omega).
\end{align}
Recall the momentum and the Green's function is defined in Eq.~\eqref{eq:pm_q_convention}.
Next, we sum over the impurity density and this fixes the external
relative momentum $\vec{q}$ to be equal to the internal momentum $\vec{q}'$. The
resulting equation is given by the following:
\begin{equation} 
\Sigma^{<}\left(\vec{p}+\frac{\vec{q}}{2},\vec{p}-\frac{\vec{q}}{2},\omega,t\right)=n_{im}\sum_{\vec{k}}T^{R}\left(\vec{p}+\frac{\vec{q}}{2},\vec{k}+\frac{\vec{q}}{2},\omega\right)G^{<}\left(\vec{k}+\frac{\vec{q}}{2},\vec{k}-\frac{\vec{q}}{2},\omega,t\right)T^{A}\left(\vec{k}-\frac{\vec{q}}{2},\vec{p}-\frac{\vec{q}}{2},\omega\right)
\end{equation}
This equation is valid for arbitrary $\vec{q}$. However, the typical situation is $|\vec{k}|,|\vec{p}|\sim k_F$ and $k_F\gg|\vec{q}|$ where $\vec{q}$ parameterized a thin-shell of scattering phase-space around the 
Fermi surface. Next, we formally expand the T-matrix as follows:
\begin{equation} \label{eq:T-matrix expansion}
T^{R}\left(\vec{p}+\frac{\vec{q}}{2},\vec{k}+\frac{\vec{q}}{2},\omega\right)=e^{\frac{1}{2}\vec{q}\cdot\vec{D}_{pk}}T_{\vec{p}\vec{k}}^{R}(\omega)\;\;,\;\;T^{A}\left(\vec{k}-\frac{\vec{q}}{2},\vec{p}-\frac{\vec{q}}{2},\omega\right)=e^{-\frac{1}{2}\vec{q}\cdot\vec{D}_{pk}}T_{\vec{k}\vec{p}}^{A}(\omega)
\end{equation}
where $\vec{D}_{pk}=\nabla_{\vec{p}}+\nabla_{\vec{k}}$ and $T_{\vec{p}\vec{k}}^{R(A)}(\omega)\equiv T^{R(A)}\left(\vec{p},\vec{k},\omega\right)$
is the T-matrix with $\vec{q}=0$. To proceed further, we substitute  Eq.~\eqref{eq:lesser_GF} and the above formula into $\Sigma^{<}$, and  and used Wigner transformation
$(\vec{q}\rightarrow-i\nabla_{\vec{r}})$ to arrive at the following expression:
\begin{align}
 & \Sigma^{<}(\vec{r},t,\vec{p},\omega)\\
= &  n_{im}\sum_{\vec{k}}T_{\vec{p}\vec{k}}^{R}(\omega)e^{-\frac{i}{2}\overleftarrow{D}_{pk}\overrightarrow{\nabla}_{r}}\left( 
-2iA(\vec{k},\omega,\vec{r},t) \hat{n}_{\vec{k}}(\vec{r},t) \right)
e^{\frac{i}{2}\overleftarrow{\nabla}_{r}\overrightarrow{D}_{pk}}T_{\vec{k}\vec{p}}^{A}(\omega)\\
= & 2\pi i n_{im}\sum_{\vec{k}}\delta(\omega-\epsilon_{k})\bigg[T_{\vec{p}\vec{k}}^{R}(\omega)\hat{n}_{\vec{k}}(\vec{r},t)T_{\vec{k}\vec{p}}^{A}(\omega)+\frac{i}{2}\bigg(T_{\vec{p}\vec{k}}^{R}(\omega)\overleftarrow{D}_{pk}\overrightarrow{\nabla}_{r}\hat{n}_{\vec{k}}(\vec{r},t)-\text{h.c.}\bigg)\bigg]+O(\nabla_{r}^{2})\label{eq:sigma_linear}
\end{align}
Lastly, we substitute Eq.~\eqref{eq:Sigmar} and \eqref{eq:sigma_linear}
into Eq.~\eqref{eq:coll-int-der} and arrive at the
following collision-integral $I^{<}(\vec{r},t,\vec{p},\epsilon_{\vec{p}})=I_{0}[n_{\vec{p}}]+I_{1}[n_{\vec{p}}]$
where, 
\begin{equation}
I_{0}[n_{\vec{p}}]=2\pi n_{im}\sum_{\vec{k}}\delta(\epsilon_{\vec{p}}-\epsilon_{\vec{k}})\bigg[T_{\vec{p}\vec{k}}^{R}\,n_{\vec{k}}\,T_{\vec{k}\vec{p}}^{A}-\frac{1}{2}\big\{ T_{\vec{p}\vec{k}}^{R}T_{\vec{k}\vec{p}}^{A},n_{\vec{p}}\big\}\bigg]\label{eq:I0-1}
\end{equation}
\begin{equation}
I_{1}[n_{\vec{p}}]=\pi n_{im}\sum_{\vec{k}}\delta(\epsilon_{\vec{p}}-\epsilon_{\vec{k}})\,i\left[T_{\vec{p}\vec{k}}^{R}\,\partial_{\vec{r}}n_{\vec{k}}\cdot\left(\vec{D}_{\vec{p}\vec{k}}T_{\vec{k}\vec{p}}^{A}\right)-\left(\vec{D}_{\vec{p}\vec{k}}T_{\vec{p}\vec{k}}^{R}\right)\cdot\partial_{\vec{r}}n_{\vec{k}}\,T_{\vec{k}\vec{p}}^{A}\right]\label{eq:I1-1-1}
\end{equation}
Eq.~\eqref{eq:I0-1} and \eqref{eq:I1-1-1} are the basic equations of the
collision integral discussed in the main text. In order to gain more
physical insight, we shall express them in terms of the scattering
$S$-matrix using the relationship between the $S$-matrix and $T$-matrix:
\begin{equation}
S_{\alpha\gamma}(\vec{p},\vec{k})=\delta_{pk}\delta_{\alpha\gamma}-2\pi i\,t_{\alpha\gamma}(\vec{p},\vec{k})\label{eq:s-t}
\end{equation}
In the above, we introduce a dimensionless on-shell T-matrix $t_{\alpha\gamma}(\vec{p},\vec{k})$:
\begin{equation}
t_{\alpha\gamma}(\vec{p},\vec{k})=\delta(\epsilon_{p}-\epsilon_{k})T_{\alpha\gamma}(\vec{p},\vec{k}).\label{eq:on-t}
\end{equation}
In the remaining of this section, we shall simply denote the retarded
T-matrix as $T_{\alpha\gamma}(\vec{p},\vec{k})=\langle\vec{p}\alpha|T(\epsilon+i\delta)|\vec{k}\beta\rangle$
and dropped the $R$ label (i.e. $T_{\alpha\gamma}(\vec{p},\vec{k})=\langle\alpha|T_{\vec{p}\vec{k}}^{R}|\beta\rangle$
in the main-text). The advanced T-matrix is related to the retarded
T-matrix by Hermitian conjugate $T_{\alpha\beta}^{A}(\vec{k},\vec{p})=T_{\beta\alpha}^{*}(\vec{p},\vec{k})$
. From unitarity of the $S$-matrix ($SS^{\dagger}=1)$, we obtained
following important relation that is sometimes called the generalized optical theorem:
\begin{equation}
2\pi\sum_{\vec{p}'}\,t_{\gamma\alpha}(\vec{p}',\vec{k})\,t_{\gamma \beta}^{*}(\vec{p}',\vec{p})
=i\, \left[t_{\alpha\beta}(\vec{p},\vec{k}) -t_{\beta\alpha}^{*}(\vec{p},\vec{k}) \right]\label{eq:opt-theorem}
\end{equation}
When we consider forward scattering $\vec{p}=\vec{k}$, this reduces to the usual optical theorem.

Let us discuss Eq.~\eqref{eq:I0-1}. For isotropic impurity, the Hermitian
part of the self-energy $\Sigma_{\vec{p}}^{H}=n_{im}(T_{\vec{p}\vec{p}}^{R}+T_{\vec{p}\vec{p}}^{A})/2$
only has charge component (i.e. proportional to identity matrix) and
it drops out in the kinetic equation $[\Sigma_{\vec{p}}^{H},\hat{n}_{p}]=0$.
In case it has a spin-component (e.g. Rashba type spin-orbit coupling),
we can combine it with $I_{0}$ and let
\begin{equation}
\mathbb{I}_{0}[n_{\vec{p}}]=-i[\Sigma_{\vec{p}}^{H},n_{\vec{p}}]+I_{0}[n_{\vec{p}}].
\end{equation}
Using Eq.\eqref{eq:opt-theorem}, we express the anti-commutator in Eq.~\eqref{eq:I0-1}
to linear order in $T$ and combined with $\Sigma_{\vec{p}}^{H}$
to arrive at the following expression:
\begin{equation}
\mathbb{I}_{0}[n_{\vec{p}}]=i\,n_{im}(n_{\vec{p}}T_{\vec{p}\vec{p}}^{A}-T_{\vec{p}\vec{p}}^{R}n_{\vec{p}})+2\pi n_{im}\sum_{\vec{k}}\delta(\epsilon_{\vec{p}}-\epsilon_{\vec{k}})T_{\vec{p}\vec{k}}^{R}\,n_{\vec{k}}\,T_{\vec{k}\vec{p}}^{A}
\end{equation}
In equilibrium, the quasiparticle is distributed according to a Fermi-Dirac
distribution $n_{F}(\epsilon_{k})$ and it does not contribute to
the collision integral $\mathbb{I}_{0}[n_{F}(\epsilon_{k})]=0$. In
the important linear response regime, the deviation of the Fermi surface
is bound on the Fermi surface:
\begin{equation}
\hat{n}_{\vec{k}}=n_{F}(\epsilon_{k})+\delta(\epsilon_{k}-\epsilon_{F})\delta n_{\vec{k}}\label{eq:linearized}
\end{equation}
where $\epsilon_{F}$ is the Fermi energy. Then, we can parametrize
$I_{0}[n_{\vec{p}}]$ with on-shell T-matrix as follows:
\begin{align}
\mathbb{I}_{0}[n_{\vec{p}}]\bigg|_{\alpha\beta} & =-n_{im}\sum_{\vec{k}}\bigg[i\delta_{\vec{k}\vec{p}}[\delta_{\beta\delta}t_{\alpha\gamma}(\vec{p},\vec{k})-\delta_{\alpha\gamma}t_{\beta\delta}^{*}(\vec{p},\vec{k})]-2\pi t_{\alpha\gamma}(\vec{p},\vec{k})t_{\beta\delta}^{*}(\vec{p},\vec{k})\bigg]\delta n_{\gamma\delta}(\vec{k}),\\
 & =-\frac{n_{im}}{2\pi}\sum_{\vec{k}}\Lambda_{\alpha\beta,\gamma\delta}(\vec{p},\vec{k})\,\delta n_{\gamma\delta}(\vec{k}),
\end{align}
where the relaxation superoperator reads,
\begin{equation}
\Lambda_{\alpha\beta,\gamma\delta}(\vec{p},\vec{k})=\delta_{\vec{p}\vec{k}}\delta_{\alpha\gamma}\delta_{\beta\delta}-S_{\alpha\gamma}(\vec{p},\vec{k})S_{\beta\delta}^{*}(\vec{p},\vec{k}).\label{eq:Lambda_super}
\end{equation}
To derive the last line, we simply spell out the product $S_{\beta\delta}^{*}(\vec{p},\vec{k})S_{\alpha\gamma}(\vec{p},\vec{k})$
using Eq.\eqref{eq:s-t}. Note the relaxation superoperator is 
positive definite and it describes the decay of the excitations on the Fermi surface (see Sec.~\ref{sec:positivity} for a proof). It is also worth noting that the conservation of charge follows from the unitarity of the $S$-matrix:
\begin{equation}
\sum_{\vec{p}}\Lambda_{\alpha\alpha,\gamma\delta}(\vec{p},\vec{k})=\delta_{\delta\gamma}-\sum_{\vec{p},\alpha}S_{\alpha\gamma}(\vec{p},\vec{k})S_{\alpha\delta}^{*}(\vec{p},\vec{k})=\delta_{\delta\gamma}-\delta_{\gamma\delta}\delta_{\vec{k}\vec{k}}=0
\end{equation}
The matrix elements of $I_{1}$ in Eq. \eqref{eq:I1-1-1} read as follows:
\begin{align}
I_{1}[n_{\vec{p}}]\bigg|_{\alpha\beta} & =\pi n_{im}\sum_{\vec{k}}\delta(\epsilon_{\vec{p}}-\epsilon_{\vec{k}})\,\left[T_{\alpha\gamma}(\vec{p},\vec{k})\,i\overleftrightarrow{D}_{\vec{p}\vec{k}}T_{\beta\delta}^{*}(\vec{p},\vec{k})\right]\cdot\nabla_{\vec{r}}n_{\vec{k},\gamma\delta}\\
 & =\pi n_{im}\sum_{\vec{k}}\vec{V}_{\alpha\beta,\gamma\delta}(\vec{p},\vec{k})\cdot\nabla_{\vec{r}}\delta n_{\vec{k},\gamma\delta}
\end{align}
where $\overleftrightarrow{D}_{\vec{p}\vec{k}}=\overrightarrow{D}_{\vec{p}\vec{k}}-\overleftarrow{D}_{\vec{p}\vec{k}}$,
$\overrightarrow{D}_{\vec{p}\vec{k}}=\nabla_{\vec{p}}+\nabla_{\vec{k}}$
and the velocity-operator defined in the above and main-text reads
as follow:
\begin{equation}
\vec{V}_{\alpha\beta,\gamma\delta}(\vec{p},\vec{k})=t_{\alpha\gamma}(\vec{p},\vec{k})\,i\overleftrightarrow{D}_{\vec{p}\vec{k}}\,t_{\beta\delta}^{*}(\vec{p},\vec{k})
\end{equation}
In deriving the second line, we used Eq.~\eqref{eq:linearized}. The
vector $\vec{V}$ is a matrix analog of side-jump defined in the degenerate
space of a given band. From the properties of the T-matrix, we can
see that the velocity operator is hermitian $\vec{V}_{\alpha\beta\gamma\delta}(\vec{p},\vec{k})=\vec{V}_{\gamma\delta\alpha\beta}^{*}(\vec{k},\vec{p})$.
To convince ourselves that the velocity matrix is simply a shift in
scattering phase, we can express $\vec{V}$ in terms of $S$-matrix
using Eq. \eqref{eq:s-t}:
\begin{equation}
I_{1}[\hat{n}_{\vec{p}}]_{\alpha\beta}=-\frac{n_{im}}{4\pi}\sum_{k}\bigg[\delta_{kp}\delta_{\alpha\gamma}\delta_{\beta\delta}2\vec{D}_{\vec{p}\vec{k}}\text{Im}S_{0}(\vec{p},\vec{k})-S_{\alpha\gamma}(\vec{p},\vec{k})i\overleftrightarrow{D}_{\vec{p}\vec{k}}S_{\beta\delta}^{*}(\vec{p},\vec{k})\bigg]\cdot\nabla_{r}\delta\hat{n}_{\vec{k},\gamma\delta}
\end{equation}
where we have used the following property $\vec{D}_{\vec{p}\vec{k}}S_{\alpha\beta}(\vec{p},\vec{k})|_{\vec{p}=\vec{k}}=\vec{D}_{\vec{p}\vec{k}}S_{0}(\vec{p},\vec{k})|_{\vec{p}=\vec{k}}\delta_{\alpha\beta}$
for isotropic impurity. While the first term is diagonal in the spin
index, the second term contributes to important spin-charge correlation.

\section{Side-jump anomalous velocity}

In this section, we discuss different approximations to the side-jump anomalous velocity that are commonly used in the literature. In the kinetic formalism we presented above, the anomalous velocity is defined by the response of the spin diagonal components of the collision intral $I_1$ to an electrochemical potential difference $\vec{\nabla}_{\vec{r}} \hat{n}_{\vec{k}}= (-\frac{\partial n_F}{\partial \epsilon})\nabla_{\vec{r}}\mu$. Substitute this expression of $\vec{\nabla}_{\vec{r}} \hat{n}_{\vec{k}}$ into $I_1$, we arrived at the following expression that defines the side-jump anomalous velocity $\vec{\omega}$:

\begin{align}
I_{1}[n_{\vec{p}}]\bigg|_{\alpha\alpha} & =2 \pi n_{im}\sum_{\vec{k},\gamma}\delta(\epsilon_{\vec{p}}-\epsilon_{\vec{k}})\,\mathrm{Im}\: \left[
\left( {D}_{\vec{p}\vec{k}}T_{\alpha\gamma}(\vec{p},\vec{k}) \right) T_{\alpha\gamma}^{*}(\vec{p},\vec{k}) \right]\cdot \left(-\frac{\partial n_F}{\partial \epsilon}\right) \nabla_{\vec{r}} \mu
 \equiv N_F \, \vec{\omega}(\vec{p},\alpha) \cdot  \nabla_{\vec{r}} \mu,
\end{align}

\begin{equation} \label{eq:AV}
    \vec{\omega}(\vec{p},\alpha) = 2 \pi n_{im}\sum_{\vec{k},\gamma}\delta(\epsilon_{\vec{p}}-\epsilon_{\vec{k}})\,\mathrm{Im}\left[
\left( {D}_{\vec{p}\vec{k}}T_{\alpha\gamma}(\vec{p},\vec{k}) \right) T_{\alpha\gamma}^{*}(\vec{p},\vec{k}) \right],
\end{equation}
\begin{equation}
 \vec{D}_{\vec{p}\vec{k}} = \nabla_{\vec{k}} + \nabla_{\vec{p}}.
\end{equation}
When Born-approximation is used, the T-matrix is replaced with the (bare) disorder potential $T_{\alpha\gamma}(\vec{p},\vec{k})=V_{\alpha\gamma}(\vec{p},\vec{k})$ and we arrived the the formula derived by Lyo and Holstein \cite{PhysRevLett.29.423}:
\begin{equation}
    \vec{\omega}^{\rm{L.H.}}(\vec{p},\alpha) = 2 \pi n_{im}\sum_{\vec{k},\gamma}\delta(\epsilon_{\vec{p}}-\epsilon_{\vec{k}})\,\mathrm{Im}\left[
\left( {D}_{\vec{p}\vec{k}}V_{\alpha\gamma}(\vec{p},\vec{k}) \right) V_{\alpha\gamma}^{*}(\vec{p},\vec{k}) \right].
\end{equation}
The generalization of the Lyo-Holstein formula was performed by Levy in Ref.~\onlinecite{levy1988extraordinary}. He used the stationary solution of a single impurity scattering problem to compute the (disorder-induced)
anomalous velocity operator $[\vec{r},V_{im}]/i$ and found:
\begin{equation} \label{eq:AV_levy}
    \vec{\omega}^{\rm{Levy}}(\vec{p},\alpha)   =2 \pi n_{im}\sum_{\vec{k},\gamma}\delta(\epsilon_{\vec{p}}-\epsilon_{\vec{k}})\,\mathrm{Im}\left[
\left( \nabla_{\vec{k}}T_{\alpha\gamma}(\vec{p},\vec{k}) \right) T_{\alpha\gamma}^{*}(\vec{p},\vec{k}) \right] 
\end{equation}
Note the difference in momentum gradient in Eq.~\eqref{eq:AV} and \eqref{eq:AV_levy}. 
While Eq.~\eqref{eq:AV_levy} is a well defined quantity, it is unclear how it should appear in the Boltzmann kinetic equation and this has generated a lot of debate cited in the main text. Our rigorous kinetic formalism provides an answer to this question unambiguously.
Indeed, Eq.~\eqref{eq:AV_levy} can also be recovered from the self-energy we described in Fig.~1 which we rewrite here for convenience (c.f.~Eq.\eqref{eq:Sigma_less}):
\begin{align}
\Sigma^{<}(\vec{p}+\frac{\vec{q}}{2},\vec{p}-\frac{\vec{q}}{2},\omega,t)
&= \frac{1}{\Omega}\sum_{\vec{r}_{a}=1}^{N_{im}}\sum_{\vec{p}'\vec{q}'}
e^{i(\vec{q}-\vec{q'})\cdot\vec{r}_{a}}
T^{R}(\vec{p}+\frac{\vec{q}}{2},\vec{p}'+\frac{\vec{q}'}{2},\omega)G^{<}(\vec{p}'+\frac{\vec{q}'}{2}, \vec{p}'-\frac{\vec{q}'}{2},\omega,t)T^{A}(\vec{p}'-\frac{\vec{q}'}{2},\vec{p}-\frac{\vec{q}}{2},\omega) 
\end{align}
If one (incorrectly) discards the factor $e^{-i\vec{q'}\cdot\vec{r}_{a}}$ and then performs the impurity average, a self-energy that is diagonal in momentum is thus obtained:
\begin{equation}
    \Sigma^{<}(\vec{p}+\frac{\vec{q}}{2},\vec{p}-\frac{\vec{q}}{2},\omega,t) \rightarrow  \Sigma^{<}(\vec{p},\omega,t)\delta_{\vec{q},0}
\end{equation}
\begin{equation}
    \Sigma^{<}(\vec{p},\omega,t)=n_{im}\sum_{\vec{p}'\vec{q}'}
T^{R}(\vec{p},\vec{p}'+\frac{\vec{q}'}{2},\omega)G^{<}(\vec{p}'+\frac{\vec{q}'}{2}, \vec{p}'-\frac{\vec{q}'}{2},\omega,t)T^{A}(\vec{p}'-\frac{\vec{q}'}{2},\vec{p},\omega) %
\end{equation}
When the $q$ dependence of the $T$-matrix is expanded using  Eq.~\eqref{eq:T-matrix expansion} and following the standard Wigner transformation described below Eq.~\eqref{eq:T-matrix expansion}, one arrives at the anomalous velocity defined in Eq.~\eqref{eq:AV_levy}. 

The important difference between Eq.~\eqref{eq:AV} and \eqref{eq:AV_levy} can be most clearly seen in the Born approximation where the T-matrix is approximated by $T_{\alpha\beta}(\vec{p},\vec{k}) = U(\vec{p}-\vec{k})( \delta_{\alpha\beta} - i \gamma (\vec{p} \times \vec{k})\cdot \sigma_{\alpha,\beta})$, then they become
\begin{align}
     \vec{\omega}(p,\alpha) &= \frac{\gamma \vec{\sigma}_{\alpha,\alpha}\times \vec{p}}{\tau_{tr}}, \\
    \vec{\omega}^{\rm{Levy}}(p,\alpha) &= \frac{\gamma \vec{\sigma}_{\alpha,\alpha}\times \vec{p}}{\tau_{qp}}. \\
\end{align}
In these formula, the numerator defines the side-jump distance and the denominator defines the typical time-scale associated with the side-jump. Here $\tau_{tr}$ and $\tau_{qp}$ are respectively the transport and quasiparticle lifetime:
\begin{align}
    \frac{1}{\tau_{tr}} &= 2\pi n_{im} \sum_{\vec{k}} |U(\vec{p}-\vec{k})|^2 \delta(\epsilon_p-\epsilon_k)(1- \vec{\hat{p}}\cdot \vec{\hat{k}}) \\
    \frac{1}{\tau_{qp}} &= 2\pi n_{im} \sum_{\vec{k}} |U(\vec{p}-\vec{k})|^2 \delta(\epsilon_p-\epsilon_k)
\end{align}
We therefore conclude that Eq.~\eqref{eq:AV} is more applicable than Eq.~\eqref{eq:AV_levy} since already at the Born approximation, $\vec{\omega}^{\rm{Levy}}(p,\sigma) $ is correct only when the impurity vertex correction can be neglected (i.e.~$\tau_{tr}=\tau_{qp}$).

\section{Solution in the diffusive limit}
\label{sec:solution}

In this section, we discuss in detail the steps to solve the kinetic equation in the maintext
\begin{equation}
   ( \partial_{t}+\vec{v}_{\vec{p}}\cdot\nabla_r ) \hat{n}_{\vec{p}}  =\hat{I}_0[\hat{n}_{\vec{p}}]+ \hat{I}_{1}[ \hat{n}_{\vec{p}}]. 
\end{equation}
The above kinetic equation is really a 2$\times$2 matrix equation that describes the dynamics of quasiparticles on a (doubly) spin-degenerate Fermi-surface (FS) in the semiclassical limit ($k_F l \gg 1$ where $l$ is the mean-free path). The quantity we would like to determine is the correction to the ground state density-matrix ($\delta \hat{n}_{\vec{p}}=  \hat{n}_{\vec{p}} -   \hat{n}_{\vec{p}}^0 $) generated by thermodynamic potentials (e.g.~electrochemical potential) that changes spatially on a scale $\gg l \gg k_F^{-1}$. This quantity satisfies the following equation of motion:
\begin{equation} \label{eq:KE_detail}
       ( \partial_{t}+\vec{v}_{\vec{p}}\cdot\nabla_r ) \, \delta \hat{n}_{\vec{p},\alpha\beta}
       = -\frac{n_{im}}{2\pi}\sum_{\vec{k}}\Lambda_{\alpha\beta,\gamma\delta}(\vec{p},\vec{k})\,\delta n_{\vec{k},\gamma\delta} 
    +
     \pi n_{im}\sum_{\vec{k}}\vec{V}_{\alpha\beta,\gamma\delta}(\vec{p},\vec{k})\cdot\nabla_{\vec{r}}\delta n_{\vec{k},\gamma\delta}
\end{equation}
When the system is close to an isotropic limit, the typical scattering rates generated by the relaxation operator $\sum_k \Lambda(p,k) \hat{k}^m$ and the typical velocity generated by the anomalous velocity operator $\sum_k \vec{V}(p,k) \hat{k}^m$ are small for large moments ($m \gg 0$). When we only retain the zeroth and first moment $m=0,1$, we are in the so-called diffusive limit and we can paramaterize $\hat{n}_{\vec{p}}$ as
\begin{equation} \label{eq:ansatz}
 N_{F}\delta\hat{n}_{\vec{p}}= 
 \rho\mathbb{1} + s_a\sigma_a +3\hat{p}_i( g_{i0}\mathbb{1}+g_{ia}\sigma_a)v_F^{-1}.
\end{equation}
The unknowns in the right hand side can be into three categories:
the charge density $\rho=N_{F}\mu_{c}$ which is a scalar, the spin-polarization
($s_{b}=N_{F}\mu_{s,b}$) and charge first-moment $g_{l0}$ which
are rank-1 tensor with 3 components, and the spin first moment $g_{lb}$
which is a rank-2 tensor with 9 independent components. It is convenient to
decompose $g_{lb}$ into three irreducible spherical tensors $G_{lb}^{m}$
with weight $m=0,1,2$. 

\begin{align} 
&G_{ja}^{m=0}=\frac{1}{3}\delta_{ja}g_{ii}
\;\;  , \; \;
G_{ja}^{m=1} =\frac{1}{2}(g_{ja}-g_{aj}) 
\;\;  , \; \;
G_{ja}^{m=2}=\frac{1}{2}(g_{ja} + g_{aj}) -\frac{1}{3}\delta_{ja}g_{ii} \\
&P_{ja}^{m=0}=\frac{1}{3}\delta_{aj}\partial_{i}s_{i}
\;\;  , \; \;
P_{ja}^{m=1}=\frac{1}{2}(\partial_{j}s_{a}-\partial_{a}s_{j}) \;\;  , \; \;
P_{ja}^{m=2}=\frac{1}{2}(\partial_{j}s_{a}+\partial_{a}s_{j})- \frac{1}{3}\delta_{aj}\partial_{i}s_{i}
\end{align}
Readers should not confuse $G_{lb}^{m}$ with
various Green functions introduced in previous sections. We have done similar decomposition for the tensor made up by taking spatial gradient along direction $j$ on the spin-density polarized along direction $a$, $\partial_j s_a$.

When we substitute Eq.~\eqref{eq:ansatz} into Eq.~\eqref{eq:KE_detail}, 
and take the zeroth ($\sum_{\vec{p}}$) and first momentum ($\sum_{\vec{p}}\vec{p}_i$), we arrive at 
a system of equations ($\vec{u}=\hat{M}\vec{u}$) that relate a vector of responses $\vec{u}=(\rho,s_{b},g_{l0},G_{lb},\partial_{l}s_{b},\partial_{l}\rho)$ with each other with a matrix $\hat{M}$ made up by the moments of the relaxation operator $\Lambda$ and anomalous velocity $\vec{V}$.
For an isotropic Fermi surface and disorder potential,
this matrix can be spanned by isotropic tensors of different rank. This allows us to write down the solution to the kinetic equation solely based
on symmetry principles. There is only one rank-2 isotropic tensor and one rank-3 isotropic tensor. They correspond to the familiar kronecker delta function $\delta_{jl}$ and the Levi-Civita function $\epsilon_{ijk}$. On the contrary, there are three 
rank-4 isotropic tensors with weight $m=0,1,2$:
\begin{equation}
\mathbb{T}_{jbla}^{m=0}=\frac{1}{3}\delta_{jb}\delta_{la}\;,\;\mathbb{T}_{jbal}^{m=1}=\frac{1}{2}\bigg[\delta_{ja}\delta_{bl}-\delta_{jl}\delta_{ba}\bigg]\;,\;\mathbb{T}_{jbal}^{m=2}=\frac{1}{2}\bigg[\delta_{ja}\delta_{bl}+\delta_{jl}\delta_{ba}\bigg]-\frac{1}{3}\delta_{jb}\delta_{la}
\end{equation}
Note they are mutually orthogonal and normalized to $2m+1$. Importantly,
they preserve the weight of the rank-two tensors, $\mathbb{T}_{jbal}^{m}G_{la}^{m}=G_{jb}^{m}$
for $m=0,1,2$.

To illustrate how the above isotropic-tensors solves the kinetic equation, let us consider
the uniform limit where Eq.~\eqref{eq:KE_detail} takes a simpler form:
\begin{equation}
\partial_{t}n_{\vec{p},\alpha\beta}=-\frac{n_{im}}{2\pi}\sum_{\vec{k}}\Lambda_{\alpha\beta,\gamma\delta}(\vec{p},\vec{k})\,\delta n_{\vec{k},\gamma\delta}, \label{eq:relaxation}
\end{equation}

Then, the zeroth moment of the above equation must take the following form:
\begin{equation}
\partial_{t}\rho=0\;,\;\partial_{t}s_{a}=-\frac{\delta_{ab}}{\tau_{s}}s_{b}
\end{equation}
The first equation is just charge conservation. Mathematically, the second equation contracts a rank-1 tensor (vector) with a rank-2 tensor to give another rank-1 tensor. Since we know physically the rank-1 tensor (spin density $s_a$) is spatially isotropic and there is only one rank-2 isotropic tensor (i.e.~$\delta_{ab}$), the spin-relaxation equation must take the above form. Using similar arguments, the solution of the first moments
must have the following form:
\begin{equation}
\partial_{t}g_{j0}=-\frac{\delta_{jl}}{\tau_{\text{tr}}}g_{l0}+\omega_{sk}\epsilon_{jka}\,G_{la}^{m=1}\label{eq:g_j0}
\end{equation}
\begin{equation}
\partial_{t}G_{jb}^{m}
=-\frac{1}{\tau_{m}}
\mathbb{T}_{jbal}^{m}G_{la}^{m}
=-\frac{1}{\tau_{m}}G_{jb}^{m}\;\;,\;\;m=0,1,2
\label{eq:gjm} 
\end{equation}
 The physical meaning of various scattering rates are transparent.
$\tau_{s}$ and $\tau_{\text{tr}}$ are the familiar spin-relaxation
time and transport lifetime (i.e.\textasciitilde momentum relaxation
time) respectively. The spin first-moment $G_{jb}^{m}$ can relax
at different rate $\tau_{m}$ according to their tensor weight. Besides
relaxation, the most interesting scattering process above is the skew-scattering
$\omega_{sk}$: it describes the ``conversion'' between charge and
spin degree of freedom. While the form of the rate equations can be
\emph{a priori} determined by symmetry, the scattering rates have to be evaluated
microscopically using the $\Lambda$ superoperator:
\begin{align}
\frac{1}{\tau_{\text{tr}}} & =\frac{n_{im}}{4\pi N_{F}}\sum_{\vec{p},\vec{k}}\vec{\hat{p}}\cdot\vec{\hat{k}}\,\Lambda_{\alpha\alpha,\gamma\gamma}(\vec{p},\vec{k}),\\
\frac{1}{\tau_{s}} & =\frac{n_{im}}{12\pi N_{F}}\sum_{\vec{p},\vec{k}}(\sigma_{a})_{\alpha\beta}\Lambda_{\beta\alpha,\gamma\delta}(\vec{p},\vec{k})(\sigma_{a})_{\gamma\delta},\\
\omega_{sk} & =\frac{n_{im}}{8\pi N_{F}}\sum_{\vec{p},\vec{k}}\Lambda_{\alpha\alpha,\gamma\delta}(\vec{p},\vec{k})\vec{\sigma}_{\gamma\delta}\cdot(\vec{\hat{k}}\times\vec{\hat{p}}),\\
\frac{1}{\tau_{m}} & =\frac{\mathbb{T}_{jbal}^{m}}{(2m+1)}\frac{3n_{im}}{2\pi N_{F}}\sum_{\vec{p},\vec{k}},\hat{p}_{j}(\sigma_{b})_{\alpha\beta}\Lambda_{\beta\alpha,\gamma\delta}(\vec{p},\vec{k})(\sigma_{a})_{\gamma\delta}\hat{k}_{l}.
\end{align}
Note that henceforth we sum over repeated indices except for the index $m$ which
is reserved for labeling the three components of the tensor, $G_{jb}$, namely $G_{jb}^{m=0,1,2}$. Furthermore, the momentum indeces ($j,l$) and spin-projection indices $(a,b)$ run over three values $x,y,z$, whereas the Greek indices $\alpha,\beta,\gamma,\delta$
are used to label elements of density matrix distribution function of an electron with momentum $\vec{p}$, $n(\vec{p})$ 
and take two possible values ($\uparrow,\downarrow$ or $0,1$). The impurity scattering  from an electron state
$\vec{p}$ to a state $\vec{k}$ is described by  $\check{\Lambda}(\vec{p},\vec{k})$.
For spinless particles, $\check{\Lambda}(\vec{p},\vec{k})$ is just
a scalar function of momentum $\vec{p}$ and $\vec{k}$. However,
for spin degenerate bands, $\check{\Lambda}(\vec{p},\vec{k})$ is
a 4th-rank tensor super-operator  acting on $\delta\hat{n}_{\vec{k}}$. Thus, for instance transport lifetime $\tau_{\text{tr}}$ is obtained by acting with $\check{\Lambda}$ on
the charge-component of the density matrix (i.e.~identity-matrix). By contrast, the spin relaxation time, $\tau_{s}$ is obtained by acting with $\check{\Lambda}$
on the spin-component of the density matrix (i.e.~Pauli-matrices).
The skew-scattering rate $\omega_{sk}$ are obtained by contracting with $\Lambda$
on a (spin) Pauli matrix and an identity matrix, as it corresponds to a spin-to-charge conversion coefficient. Lastly, $\tau_{m}$ describes
relaxation time of various tensor-rank spin-current distributions that are obtained by contracting $\check{\Lambda}$
with the tensors $\mathbb{T}^{m}$.

In the presence of spatial non-uniformity ($\partial_{l}\rho$, $\partial_{l}s_{b}$),
the collision integral $\hat{I}_{1}$ is non-vanishing and it has
two important consequences: it renormalizes the velocity of charge
and spin flow and introduced additional spin-charge coupling that
is responsible for the side-jump mechanism. For isotropic disorder
and Fermi surface, we can also write down equations of motion on
symmetry grounds using five isotropic tensors $\delta_{jl}$, $\epsilon_{jka}$
and $\mathbb{T}_{jkba}^{m}$. The results are

\begin{align}
 \partial_{t}\rho+\partial_{j}\mathbb{J}_{j}&=0, \\
 \partial_{t}s_{b}+\partial_{j}\mathbb{J}_{jb}&=- s_{b}/\tau_{s}, \\
\mathbb{J}_{j}  &=(1-\Omega_{c}) \, g_{j0} -  \Omega_{cs}\, \epsilon_{jka} \, G_{ka}^{m=1},  \\
\mathbb{J}_{jb}  &=\sum_{m=0}^2(1-\Omega_{m}) \, G_{jb}^{m}+ \Omega_{sc} \, \epsilon_{jbk}\, g_{k0}\\
 D \begin{pmatrix}\Omega_{c} -1& \Omega_{sc}\\
-2\Omega_{cs} & \Omega_{1}-1
\end{pmatrix}\begin{pmatrix}\partial_{i}\rho\\
P_{i}
\end{pmatrix} 
&=\begin{pmatrix}
1 & -\theta_{sk} \\
2\theta_{sk} & \gamma_{1}
\end{pmatrix}\begin{pmatrix} g_{i0}\\
G_{i}
\end{pmatrix}  \\ \label{G-m}
 D(1-\Omega_{m=0})P_{ib}^{m=0}  &=- \gamma_{m=0}\, G_{ib}^{m=0}  \\
 D(1-\Omega_{m=2})P_{ib}^{m=2}  &=- \gamma_{m=2}\, G_{ib}^{m=2}
\end{align}
Besides $\tau_{\text{tr}},\text{\ensuremath{\tau_{s}},\ensuremath{\tau_{m}}}$
and $\omega_{sk}$,  the above equations also depend on a number of  additional
(dimensionless) kinetic coefficients arising from the super-operator $\hat{\vec{V}}$, which in turn is related to the gradient correction to the collision integral $I_1[n_{\vec{p}}]$. They are given by the following expressions:
\begin{align}
\Omega_{c}= & \frac{\pi n_{im}}{2N_{F}\epsilon_F}\sum_{\vec{p},\vec{k}}\vec{V}_{\alpha\beta,\beta\alpha}(\vec{p},\vec{k})\cdot\vec{k},\\
\Omega_{cs}= & \frac{\text{\ensuremath{\pi n_{im}}}}{4N_{F}\epsilon_F}\sum_{\vec{p},\vec{k}}\vec{k}\times\vec{V}_{\gamma\gamma,\alpha\beta}(\vec{p},\vec{k})\cdot\vec{\sigma}_{\alpha\beta},\\
\Omega_{sc}= & \frac{\text{\ensuremath{\pi n_{im}}}}{4N_{F}\epsilon_F}\sum_{\vec{p},\vec{k}}\vec{\sigma}_{\alpha\beta}\cdot\vec{k}\times\vec{V}_{\beta\alpha,\gamma\gamma}(\vec{p},\vec{k}),\\
\Omega_{m}= & \frac{\mathbb{T}_{jbal}^{m}}{(2m+1)}\frac{\pi n_{im}}{2N_{F}\epsilon_F}\sum_{\vec{p},\vec{k}}p_{j}(\sigma_{b})_{\alpha\beta}\big(V_{\beta\alpha,\gamma\delta}\big)_{l}(\sigma_{a})_{\gamma\delta}.
\end{align}
Here $\Omega_{c}$ and $\Omega_{m}$ are the renormalization to
the charge flow and spin flow velocity. Besides velocity renormalization,
$\Omega_{cs}$ is the spin to charge coupling and $\Omega_{sc}$ is
the charge to spin coupling and $\gamma_m = \tau_{\text{tr}}/\tau_m$ where $m=0,1,2$.

Next, we provide the explicit expressions relating the different kinetic coefficients in the diffusion equation to the parameters of a single-impurity T-matrix. For rotational and parity symmetric disorder,
the T-matrix can be paramatereized by two scalar functions $A$ and
$B$ in the following form:
\begin{equation}
T(\vec{p},\vec{k},\epsilon)=A(\vec{p}\cdot\vec{k},|\vec{p}|,|\vec{k}|,\epsilon)+i\,B(\vec{p}\cdot\vec{k},|\vec{p}|,|\vec{k}|,\epsilon)\,(\vec{p}\times\vec{k})\cdot\boldsymbol{\sigma}\,k_F^{-2}
\end{equation}
where we allow $|\vec{p}|\ne|\vec{k}|$. For notational simplicity,
we shall not show the arguments of $A$ and $B$ explicitly. Substituting
the above parametrization into the general formula for scattering
rates defined in $\Lambda$ and $\vec{V}$, we obtain at the following
expressions\textbf{}:
\begin{align}
\frac{1}{2\pi N_{F}n_{im}\tau_{\text{tr}}} & =\int\frac{d\Omega}{4\pi}(|A|^{2}+|B|^{2} \sin^{2}\theta)(1-\cos\theta)>0\\
\frac{1}{2\pi N_{F}n_{im}\tau_{s}} & =\int\frac{d\Omega}{4\pi}|B|^{2} \sin^{2}\theta>0\\
\frac{\omega_{sk}}{2\pi N_{F}n_{im}} & =\int\frac{d\Omega}{4\pi} \text{Im\,}AB^{*}\sin^{2}\theta\\
\frac{\omega_{sw}}{2\pi N_{F}n_{im}} & =\int\frac{d\Omega}{4\pi} \text{Re\,}AB^{*}\sin^{2}\theta\\
\frac{\Omega_{c}}{2\pi N_{F}n_{im} \epsilon_F^{-1}} & =\int\frac{d\Omega}{4\pi}\text{Im}(A'A^{*}) (1+\cos\theta)+\text{Im}(B'B^{*})\sin^{2}\theta(1+\cos\theta)\\
\frac{\Omega_{cs}}{2\pi N_{F}n_{im} \epsilon_F^{-1}} & =\int\frac{d\Omega}{4\pi}\,\text{Re}(AB^{*}) (1-\cos\theta)+\frac{1}{2}\bigg[\text{Re}(AB'^{*}-A'B^{*})+|B|^{2}\bigg] \sin^{2}\theta\\
\frac{\Omega_{sc}}{2\pi N_{F}n_{im} \epsilon_F^{-1}} & =\int\frac{d\Omega}{4\pi}\,\text{Re}(AB^{*}) (1-\cos\theta)+\frac{1}{2}\bigg[\text{Re}(AB'^{*}-A'B^{*})-|B|^{2}\bigg] \sin^{2}\theta\\
\frac{\Omega_{m=0}}{2\pi N_{F}n_{im} \epsilon_F^{-1}} & =\int\frac{d\Omega}{4\pi}\,\text{Im}(A'A^{*}) (1+\cos\theta)-\text{Im}(B'B^{*}) \sin^{2}\theta(1+\cos\theta)\nonumber \\
 & \;\;-2\bigg[\text{Im}(AB^{*}) (1-\cos\theta)+\frac{1 }{2}\text{Im}(AB'^{*}-A'B^{*})\sin^{2}\theta\bigg]\\
\frac{\Omega_{m=1}}{2\pi N_{F}n_{im}  \epsilon_F^{-1} } & =\int\frac{d\Omega}{4\pi}\,\text{Im}(A'A^{*}) (1+\cos\theta)\nonumber \\
 & \;\;-\bigg[\text{Im}(AB^{*}) (1-\cos\theta)+\frac{1 }{2}\text{Im}(AB'^{*}-A'B^{*})\sin^{2}\theta\bigg]\\
\frac{\Omega_{m=2}}{2\pi N_{F}n_{im} \epsilon_F^{-1}} & =\int\frac{d\Omega}{4\pi}\,\text{Im}(A'A^{*}) (1+\cos\theta)-\frac{2}{5}\text{Im}B'B^{*}\sin^{2}\theta(1+\cos\theta)\nonumber \\
 & \;\;+\bigg[\text{Im}(AB^{*}) (1-\cos\theta)+\frac{1 }{2}\text{Im}(AB'^{*}-A'B^{*})\sin^{2}\theta\bigg]
\end{align}
where the derivatives read:
\begin{align}
A' & =\frac{1}{2}\frac{1}{k_F}\frac{dA}{d|p|}\bigg|_{p=k_F}+\frac{1}{2}\frac{1}{k_F}\frac{dA}{d|k|}\bigg|_{k=k_F}+\frac{dA}{d(\vec{p}\cdot\vec{k})}\bigg|_{p=k_F}\\
B' & =\frac{1}{2}\frac{1}{k_F}\frac{dB}{d|p|}\bigg|_{p=k_F}+\frac{1}{2}\frac{1}{k_F}\frac{dB}{d|k|}\bigg|_{k=k_F}+\frac{dB}{d(\vec{p}\cdot\vec{k})}\bigg|_{p=k_F}
\end{align}
and 
\begin{equation}
\gamma_{0}=\frac{\tau_{\text{tr}}}{\tau_{0}}=1-2\omega_{sw}\tau_{\text{tr}}\;\;,\;\;\gamma_{1}=\frac{\tau_{\text{tr}}}{\tau_{1}}=1-\omega_{sw}\tau_{\text{tr}}\;\;,\;\;\gamma_{2}=\frac{\tau_{\text{tr}}}{\tau_{2}}=1+\omega_{sw}\tau_{\text{tr}}
\end{equation}

In the next section, we show that for a magnetic impurity in the Kondo regime with a doublet ground state, the functions that parametrize the T-matrix take the following
form (c.f.~Eq.~\eqref{eq:t-matrix_final}):
\begin{align} 
A &= -\frac{e^{i\eta_{0}} \sin \eta_{0}}{\pi N_F} - \left[ \frac{e^{i\eta_{1}} \sin\eta_{1}}{\pi N_F k_F^2}  - \frac{2 e^{i\eta_{2}} \sin\eta_{2}}{\pi N_F k_F^2} \right]  \vec{p}\cdot \vec{k} \\
B &= -\frac{e^{i\eta_{1}}}{\pi N_F} \sin\eta_{1} + \frac{e^{i\eta_{2}}}{\pi N_F}  \sin\eta_{2} \\
A' &= - \frac{e^{i\eta_{1}}}{\pi N_F k^2_F}  \sin\eta_{1} - 
\frac{2 e^{i\eta_{2}}}{\pi N_F k^2_F}  \sin\eta_{2 }\\
\end{align}
and $B'=0$. These impurity parameters lead to the following kinetic coefficients associated with $I_0$:\,
\begin{align}
\frac{1}{ \tau_{\text{tr} } } &= \frac{4 \epsilon_F n_{im}}{9\pi n_c } \left[9-\cos 2 (\eta_0-\eta_1) - 2 \cos 2 (\eta_0-\eta_2)- \cos 2 \eta_1 - 4 \cos  2 \eta_2\right],\\
\frac{1}{\tau_{s}} &= \frac{16 \epsilon_F n_{im}}{9\pi n_c} \sin ^2(\eta_1-\eta_2),\\
\omega_{sk} &= 
\frac{16\epsilon_F n_{im} }{9\pi n_c}
\sin \left(\eta_0\right) \sin \left(\eta _0-\eta _1-\eta_2\right) \sin \left(\eta_1-\eta_2\right),\\
\omega_{sw} &= 
\frac{16 \epsilon_F n_{im}}{9 \pi n_c }
\sin \left(\eta_0\right)
\cos \left(\eta_0-\eta _1-\eta_2\right)
\sin \left(\eta_1-\eta_2\right)
\end{align}
and the following (dimensionless) coefficients associated with $I_1$ :
\begin{align}
\Omega_c &= 
\frac{4 n_{im}}{3\pi n_c} \sin \left(\eta _0\right) \left[3 \cos \left(\eta _0\right)-\cos \left(\eta _0-2 \eta_1\right)-2 \cos \left(\eta _0-2 \eta _2\right)\right],\\
\Omega_{cs} &= 
\frac{2 n_{im}}{9\pi n_{c}} \left[4+ 3 \cos 2 \left(\eta _0-\eta_1\right)+3 \cos 2 \eta_1-3
   \cos 2 \left(\eta _0-\eta _2\right)-4 \cos 2 \left(\eta _1-\eta _2\right)-3
   \cos 2 \eta _2\right],\\
  \Omega_{sc} &= 
  \frac{4 n_{im}}{3\pi n_c} \cos \eta _0 \left[\cos \left(\eta _0-2 
  \eta_1\right)-\cos \left(\eta_0 -
  2 \eta _2\right)\right],\\
  \Omega_{m=0} &= 
  \frac{4 n_{im}}{3\pi} \sin \left(\eta _1\right) \left[3 \cos \left(2 \eta _0-\eta _1\right)+\cos \left(\eta
   _1\right)-4 \cos \left(\eta _1-2 \eta _2\right)\right],\\
   \Omega_{m=1} &= 
   \frac{2 n_{im}}{3\pi n_c} \left[3 \sin 2 \eta _0-2 \sin 2 \left(\eta_0-\eta_1\right)-\sin 2 \left(\eta _0-\eta_2\right)-2 \sin 2 \left(\eta _1-\eta
   _2\right)-3 \sin 2 
   \eta _2\right],\\
   \Omega_{m=2} &=
   \frac{4 n_{im}}{3\pi n_c} \sin \left(\eta _2\right) \left[3 \cos \left(2 \eta _0-\eta _2\right)-2 \cos \left(2
   \eta _1-\eta _2\right)-\cos \eta _2\right].
\end{align}

In the above expressions, $n_c = k^3_F/3 \pi^2 = 4N_F \epsilon_F/3$ is the density  of carriers in the conduction band, where $\epsilon_{F}=k_F^{2}/2m^*$ is the Fermi energy and $N_{F}=m^* k_F/(2\pi^{2})$ the density of states at the Fermi level.  

\section{quantum impurity model}\label{sec:QIM}
 \subsection{Model Hamiltonian}
   The kinetic theory developed in the main text and in previous section allows to treat resonant scattering in the presence of strong SOC, and thus we  apply it to a metal contaminated by a dilute ensemble of randomly distributed magnetic impurities in the Kondo regime. In particular, the model introduced below describes rare earth Cerium impurities in  alloys such as Ce$_x$La$_{1-x}$Cu$_6$. Copper is a low-atomic number metal with negligible SOC in its band structure. The host alloy LaCu$_6$ to which Ce is added,  is a non-magnetic alloy which forms a regular structure with crystal group Pnma. 
The bands of this alloy near the Fermi energy retain a large weight on the Cu 4s, and the La 6s and 5d orbitals, which are rather extended and therefore should have a very small spin-orbit splittings. The 4f-orbital of La, which is very compact and should have large spin-orbit is empty in the atomic limit and  leads to a very narrow empty band that, according to relativistic density functional calculations, is found at $\gtrsim 2$ eV above the Fermi energy
\url{https://materialsproject.org/materials/mp-636256/}. Since DFT tends to underestimate this type of gaps, we expect this orbital to have a small effect on the spin orbit splitting of the bands near the Fermi energy. Therefore, we expect also a negligible intrinsic contribution to the spin Hall effect in LaCu$_6$.
   Therefore, the extrinsic effects in the spin-charge transport are expected to dominant. The extrinsic effects are caused by scattering of the conduction electrons with Cerium impurities. The latter 
 contain a single electron in its $f$ shell (4f$^1$ 5s$^2$) that experiences a strong SOC ($\sim 100$ meV~\cite{Kawakami1986,HewsonKondo,NewnsRead_JPhysC_1983}).  Lanthanum (5d$^1$ 6s$^2$) is also a rare earth but has no electrons in the 4f-shell. It merely allows to substitute Ce so that it becomes possible to study alloys that interpolate between the crystalline ``Kondo lattice'' alloy Ce$_2$Cu$_6$ (i.e. for $x = 2$) and the dilute alloy limit where $x\ll 1$~\cite{Sumiyama1986}.

 The strong SOC in the f-shell of Ce  splits the 4f level into two multiplets with $j = 5/2$ and $j=7/2$, being the $j=5/2$ the one with the lowest energy. The higher energy multiplet plays no role in the low-temperature transport properties of the alloy. In addition, crystal fields arising from the lattice environment of the Ce impurity 
 further split the $j=5/2$ multiplet into a doublet ($
 \Gamma_7$) and a quartet ($\Gamma_8$) separated by an energy $\sim 10$ meV or $\sim 100$ K. For the Ce$_x$La$_{1-x}$Cu$_6$ system the doublet,  $\Gamma_7$, is the ground state~\cite{Kawakami1986,HewsonKondo}. Strictly speaking, crystal field effects break rotational invariance, but since in a uniform system translational invariance is restored by averaging over a random impurity distribution, we expect rotational invariance to be restored by the disorder average in a polycrystalline sample and therefore a rotational invariant model to provide a reasonably good description of transport in dilute Cerium alloys~\cite{Costi_prl_2000}. Thus, in order to model in a simple way the low-lying doublet/quartet structure, we treat the impurity orbital as a singly occupied $l= 1$ $p$ orbital with a strong $\vec{l}\cdot \vec{s}$ type SO ($\vec{l}$ being the angular momentum and $\vec{s}$ the spin of the f-electron), which splits the level into a $j=1/2$ doublet and a $j=3/2$ quartet.  Besides capturing the degeneracy of the ground state (which is important for the Kondo Physics, as explained below) the impurity Hamiltonian is fully invariant under rotations generated by $\vec{j} = \vec{l}+  \vec{s}$, which is instrumental for the analytical solution of the Boltzmann equation.
 
  In addition to SOC, when two (or more) electrons occupy the ground state multiplet, they experience a strong Coulomb repulsion $U$. This correlation effect is responsible for the Kondo effect (with a Kondo temperature of $\sim 1$ K~\cite{Kawakami1986,Sumiyama1986,HewsonKondo,Costi_prl_2000}) that is observed as a  minimum of the resistivity followed by a saturation as the temperature tends to zero~\cite{Sumiyama1986,HewsonKondo}. In the lattice limit, i.e. for $x\to 2$, further anomalies are observed as a consequence of the formation of a \emph{heavy} fermion bands. Note that these alloys do not become magnetic at low temperature even for high concentration of Ce impurities, which is a consequence of the enhancement of the Kondo temperature relative to the magnetic ordering temperature resulting from the large orbital degeneracy arising from the $f$ orbitals~\cite{HewsonKondo,Kawakami1986}. In order to describe the impurity embedded in the metallic host, we use the following extension of the Anderson impurity model~\cite{Anderson_PhysRev.124.41,Costi_JPhysC_1994}.
\begin{align}
H = \sum_{\vec{k},\alpha} \epsilon_{\vec{k}} \: c^{\dag}_{\vec{k}\alpha} c_{\vec{k}\alpha} +
\epsilon_{0} \sum_{m=\pm} a^{\dag}_{m} a_{m} + \sum_{k,m=\pm}  \left[ V_{k} c^{\dag}_{k m}  a_{m} + V^{*}_{k} a^{\dag}_{m} c_{k m}\right] 
+ \tfrac{1}{2}U \sum_{m=\pm} a^{\dag}_{m} a^{\dag}_{-m} a_{-m} a_{m} + H^a_{\mathrm{other}}, \label{eq:ham1}\\
\end{align}
In the rotationally-invariant model of Ce impurities described above, 
the ground state doublet
 is hybridized wit the $j=\tfrac{1}{2}, l=1, s =\tfrac{1}{2}$ channel of conduction electrons, which are described by the following set of anti-commuting operators:
\begin{equation}
c^{\dag}_{k m}   = \sum_{\alpha} \int \frac{d\vec{\hat{k}} }{4\pi} \,  
c^{\dag}_{\vec{k} \alpha}  F^{l=1,j=1/2}_{\alpha,m}(\vec{\hat{k}}),\label{eq:c11}
\end{equation}
where 
\begin{equation}
F^{l=1,j=1/2}_{\alpha,m}(\vec{\hat{k}}) = \frac{\left( \vec{\sigma}\cdot 
\vec{\hat{k}} \right)_{\alpha,m}}{\sqrt{4\pi}}, \label{eq:c12}
\end{equation}
which is obtained from the spinor spherical Harmonics for the $j=\tfrac{1}{2}$ doublet
originating from the $l = 1$ and $s= \tfrac{1}{2}$ scattering states. Indeed, since $\epsilon_{\vec{k}} = \tfrac{\hbar^2 k^2}{2m^*} = \epsilon_{k}$, it is  also possible to write the kinetic energy operator
of the conduction electrons in terms of the partial waves operators as follows
\begin{equation}
H^c_0 = \sum_{k,m=\pm} \epsilon_{k} \: c^{\dag}_{k m} c_{k m} + H^c_{\mathrm{other}}. \label{eq:ham0c}
\end{equation}
The second term describes the kinetic energy of other conduction-electron scattering channels.  In addition, 
the quantum impurity  contains other orbitals/multiplets that couple to those additional channels and are  described by the term $H_{\mathrm{other}}$ in Eq.~\eqref{eq:ham1}. Neglecting many-body effects, their Hamiltonian reads~\cite{Costi_JPhysC_1994}
 \begin{align}
H^a_{\mathrm{other}} = \sideset{}{^{\prime}}\sum_{j} \sum_{m=-j}^{+j} \left\{   \epsilon^{j}_{0} a^{\dag}_{jm} a_{jm} + \sideset{}{^{\prime}}\sum_{k} \left[  V^{j}_{k} c^{\dag}_{kjm} a_{jm} + \left( V^{j}_{k}\right)^* a^{\dag}_{jm} c_{kjm} \right]  \right\}. 
 \end{align}
In this expression the prime in the sums over $j$ means that we need to exclude the multiplet with $j =\tfrac{1}{2}, l =1, s =\tfrac{1}{2}$ which is described by first four terms in the Hamiltonian of Eq.~\eqref{eq:ham1}. In the next  subsection, when computing the T-matrix, we consider a simplified version of $H_{\mathrm{other}}$ that accounts for the two additional that are closest levels in energy to the doublet with $j=1/2,l=1,s=1/2$. One of the levels is the $s$-orbital with $j=1/2, l = 0$ and $s =\tfrac{1}{2}$ and the other channel is the higher energy quartet with $j=3/2$, $l=1$ and $s =\tfrac{1}{2}$. Thus,
\begin{align}
H_{\mathrm{other}} = \sum_{j=1/2,3/2}
 \sum_{m=-j}^{+j}  \tilde{\epsilon}^{j}_0 \tilde{a}^{\dag}_{jm}\tilde{a}_{jm} +
\sum_{k,m} \left[ \tilde{V}^{j}_{k} c^{\dag}_{j km} \tilde{a}_{j m} +  \left(\tilde{V}^j_{k} \right)^* \tilde{a}^{\dag}_{j m} c_{j km}  \right], \label{eq:hamother} 
\end{align}
where $\tilde{a}_{j m}$ ($\tilde{c}^{\dag}_{j km}$) creates an electron in impurity (conduction band) with
  $j=1/2,3/2$ and $m 
  in \{-j,  \ldots, j\}$. For example, for $j=1/2$ (and $l=0, s=1/2$) we have:
\begin{equation}
\tilde{c}^{\dag}_{1/2 k m}   =  \sum_{\alpha}\int \frac{d\vec{\hat{k}} }{4\pi} \,  
c^{\dag}_{\vec{k} \alpha}  F^{l=0,j=1/2}_{\alpha,m}(\vec{\hat{k}}) \label{eq:c01}
\end{equation}
The expressions for $F^{l=1,j=3/2}{
\alpha m}(\hat{k})$
where 
\begin{equation}
F^{l=0,j=1/2}_{\alpha,m}(\vec{\hat{k}}) = \frac{\mathbb{1}}{\sqrt{4\pi}},\label{eq:c02}
\end{equation}
which is obtained using the  Clebsch-Gordan coefficients for $j=\tfrac{1}{2}, l =0$ and $s=\tfrac{1}{2}$. Similarly, we can obtain  expressions for $F^{l=1,j=3/2}_{\alpha,m}$, but  the expressions are a too long to be reproduced here. 

\subsection{Local moment regime and the Kondo Hamiltonian}\label{subsec:localmom}

 Ignoring for a moment the additional orbitals and channels described by $H^c_{\mathrm{other}} + H^a_{\mathrm{other}}$,
in a restricted Hartree-Fock mean-field approach~\cite{Anderson_PhysRev.124.41}, the interaction term in Eq.~\eqref{eq:ham1} can be approximated as follows:
 \begin{equation}
H_U = \tfrac{1}{2}U \sum_{m=\pm} a^{\dag}_{m} a^{\dag}_{-m} a_{-m} a_{m}
= U n_{a+} n_{a-} \simeq U\left(  \langle n_{a-}\rangle \: n_{a+ } + \langle n_{a+}\rangle \: n_{a-} \right)  - U \langle  n_{a-}\rangle  \langle n_{a+} \rangle. 
 \end{equation}
 By self-consistently determining the occupations $\langle n_{a\pm}\rangle$,  
 solutions are found~\cite{Anderson_PhysRev.124.41,HewsonKondo} for which $\langle n_{a+}\rangle \neq \langle n_{a-} \rangle$. This means 
  the impurity orbital develops a finite magnetization, i.e.
 \begin{equation}
\langle S^a_z \rangle = \tfrac{1}{2} \left[
\langle n_{a +}\rangle - \langle n_{a-}\rangle \right] \neq 0.
 \end{equation}
 This type of solutions are also captured by more sophisticated mean field approaches such like the  LDA+U (see e.g. Refs.~\cite{guo2009enhanced,Shick_PhysRevB.84.113112}).
 Note that, unlike the original Anderson model, for the present model the pseudo-spin operator $S^a_z$ does not correspond to the projection on the $z$-axis of the orbital spin $\vec{s}$ but to the projection of the total angular momentum $\vec{j}$. Below, we shall
 often refer to this spin as pseudo-spin in order to avoid confusing it with the actual impurity orbital spin $\vec{s}$.
 
 The unrestricted Hartree-Fock approach describes the formation of a local moment.  However, Anderson's mean field approach~\cite{Anderson_PhysRev.124.41}  as well as other more sophisticated approaches such as LDA+ U~\cite{guo2009enhanced,Shick_PhysRevB.84.113112} fail to capture (pseudo-) spin flip scattering processes where electrons hop via virtual transitions on and off  the localized orbital and modify its (pseudo-) spin orientation.  
 
 In order to make progress with the description of  spin-flip processes, let us first consider the model in the limit where  $V_{k} = 0$. In this limit, the orbital occupation $n_a = \sum_{m=\pm} n_{am}$ commutes with the Hamiltonian. Thus, the Hilbert of   the model for $V_{k} = 0$ spits into a direct sum of three subspaces where the impurity orbital occupation take (eigen) values $n_a = 0, 1, 2$. 
The ground state energy within each subspace for $N$ electrons is $E^{n_a}_0$. In the Kondo regime of the above quantum impurity model,  the ground state of the $n_a = 1$ sector is the absolute (degenerate) ground state, which means that
\begin{equation}
E^{n_{a} = 1}_0 < E^{n_a=0}_0 \leq E^{n_a=2}_0
\end{equation} 
Hence,
\begin{equation}
0 < E^{n_a=0}_0-E^{n_{a}=1}_0 \leq E^{n_a=2}_0-E^{n_a=1}_0
\end{equation} 
The energy differences between the different ground states are
\begin{align}
E^{n_a=0}_{0} - E^{n_a=1}_{0} &= (2\epsilon_F + E^{N-2}_{0F}) - 
\left( \epsilon_0 + \epsilon_F + E^{N-2}_{0F}\right) = \epsilon_F -\epsilon_0,\\
E^{n_a=2}_0 - E^{n_a=1}_{0} &= \left( 2\epsilon_0 + U + \epsilon_F + E^{N-2}_{0F}\right) 
- \left( \epsilon_0 + \epsilon_F + E^{N-2}_{0F}\right) =   
\epsilon_0 + U -\epsilon_F.
\end{align}
Here $E^{N-2}_{0F} =  \sum_{k_j\leq k_F}^{N-2} \epsilon_{k_j}$ is the ground state energy of 
the conduction electrons in the channel with $j=\tfrac{1}{2}, l=1, s=\tfrac{1}{2}$ with $N-2$ electrons. 
Thus, in the Kondo regime we must have that 
\begin{equation}
0 < \epsilon_F - \epsilon_0  < \epsilon_0 + U -\epsilon_F  \Rightarrow
 \epsilon_0 < \epsilon_F  \qquad \epsilon_0 + U > \epsilon_F
\end{equation}
Thus, a local moment  appears when   the energy of orbital, $\epsilon_0$ is  below the Fermi level $\epsilon_F$, but the energy cost to add a second electron to the singly occupied orbital, i.e. $\epsilon_0 + U $, is larger than the Fermi energy. In addition, when $V_k$ is switched on, the ground state of the $n_a=1$ sector will remain the ground state provided the linewidth  $\Delta = \pi N_F |V_{k=k_F}|^2$ is much smaller than than  the separation between these two states, i.e. $(\epsilon_0 + U  - \epsilon_0 =  U$). Under such conditions, the orbital occupation $\langle n_a \rangle \approx 1$ and a local moment exists, that is, the impurity becomes magnetic. 
  
  Note that the ground state of the $n_a = 1$ subspace is double degenerate corresponding to the two possible values of $m =\pm \tfrac{1}{2}$ of the electron in the impurity orbital. This degeneracy is lifted by the virtual transitions that become possible for $V_k \neq 0$ thus allowing conduction electrons hop on and off the orbital. In order to obtain the effective Hamiltonian that describes the result of such virtual transitions (scattering processes) and the lifting of the ground state degeneracy of the ground state in the $n_a =1$ subspace, we apply canonical
  transformation to the original Anderson model in Eq.~\eqref{eq:ham1} in order to eliminate to leading
  order the hybridization with the conduction band described by
 \begin{equation}
    H_V =  \sum_{k,m=\pm}  \left[ V_{k} c^{\dag}_{k m}  a_{m} + V^{*}_{k} a^{\dag}_{m} c_{k m}\right] 
 \end{equation}
The resulting Hamiltonian is then projected onto the subspace with  $n_a = 1$. Mathematically, 
\begin{align}
H^{\prime} = e^{-S} H e^{S}   = H_0 +  \left(H_V  - \left[S, H_0\right] \right)   + \left( \frac{1}{2}  \left[S,\left[S, H_0\right]\right] - \left[ S, H_V \right] \right) + O(V^3)
\end{align}
Thus, to leading order in $H_V$  we  require that the
the operator $S\sim O(V)$ eliminates the $H_V$:
\begin{equation}
H_V  - \left[ S, H_0 \right] = 0 \Rightarrow \left[S, H_0 \right]  = H_V. \label{eq:cond}
\end{equation}
With this choice, the transformed Hamiltonian becomes:
\begin{equation}
H^{\prime} =  
H_0 + \frac{1}{2} \left[ H_V, S\right] + O(V^3).
\end{equation}
The solution to Eq. ~\eqref{eq:cond} can be written as follows:
\begin{align}
S &= \int \frac{d\epsilon}{2\pi i} G^{+}_0(\epsilon) H_V G^{-}_0(\epsilon), \qquad
G^{\pm}(\epsilon) = \frac{1}{\epsilon - H_0 \pm i 0^{+}}.
\end{align}
In particular, we are interested on the projection of $H^{\prime}$ on the subspace where $n_a = 1$. 
Let $\mathcal{P}_1$ be the projection operator on such subspace.
In addition, we shall  write $H_V = \mathcal{V} + \mathcal{V}^{\dag}$, where $ \mathcal{V} = \sum_{k,m=\pm} V_k c^{\dag}_{km} a_{m}$. Due to the presence of the projectors $\mathcal{P}_1$, terms containing two powers of $\mathcal{V}$
or $\mathcal{V}^{\dag}$ vanish. Dropping a constant term, $\epsilon_0$,
we are left with
\begin{align}
H_K = H^c_0 + \mathrm{Im}\:  \int \frac{d\epsilon}{2\pi}  \mathcal{P}_1 \left[  \mathcal{V} G^{+}_0(\epsilon) \mathcal{V}^{\dag} G^{-}_0(\epsilon) +  \mathcal{V}^{\dag} G^{+}_0(\epsilon) \mathcal{V} G^{-}_0(\epsilon)    \right]  \mathcal{P}_1 + O(V^3)
\end{align}
The first of the two terms in the right hand-side describes virtual transitions from the single occupied subspace to the doubly occupied subspace and back to the singly occupied subspace. The second one describes virtual transitions from the singly occupied subspace to the subspace where the orbital is empty.  By neglecting the momentum dependence of $G^{\pm}_0(\epsilon)$ and approximating $V_{k} \simeq  V_{k=k_F} = V$, we arrive at the Kondo  Hamiltonian:
\begin{align}
H_{K} &= H^c_0 + H_{ac}\label{eq:hamac1},\\
H^c_0 &= \sum_{k,m} \epsilon_{k}\: c^{\dag}_{km} c_{k m} \\
H_{ac} &= -\frac{J}{2} \sum_{k,k^{\prime},m,m^{\prime}}  \mathcal{P}_1    a^{\dag}_{m} c_{k m}   c^{\dag}_{k^{\prime}\label{} m^{\prime}}  a_{m^{\prime}} \mathcal{P}_1 + \tilde{U} \sum_{k,k^{\prime},m} c^{\dag}_{k m} c_{k^{\prime} m}, \label{eq:hamac2}\\
J &= 2 |V|^2 \left[ \frac{1}{\epsilon_0 + U - \epsilon_F} + \frac{1}{\epsilon_F - \epsilon_0} \right],\\
\tilde{U}& = - \frac{|V|^2}{\epsilon_0 + U - \epsilon_F} 
\end{align}
This form of the projected Hamiltonian is particularly suited for the mean field treatment to be described in the following subsection, but it is neither the most commonly encountered in the literature not particularly illuminating from the physical point of view.  To clarify its physical significance, we introduce the spin operators
\begin{align}
S^a_{m,m^{\prime}} &= a^{\dag}_{m} a_{m^{\prime}} - \frac{\delta_{m,m^{\prime}}}{2} 
n^a,\\
n^a &= \sum_{m=\pm} a^{\dag}_m a_m,\\
S^a_{+} &= S^a_{+-} \quad S^a_{-} = S^{a}_{--} \quad S^a_z = S^{a}_{++} = - S^a_{--}.
\end{align}
and similar definitions for $\vec{S}^c(0)$ with the replacements $a_m\to c_m(0)$, etc. 
Therefore, 
\begin{align}
H_{ac} &=  \frac{J}{2} \sum_{m,m^{\prime}}   S^a_{m,m^{\prime}}  S^c_{m^{\prime},m}(0)  + U n^c(0)
 =   \frac{J}{2} \mathcal{P}_1 \left[ S^c_{+}(0)  S^a_{-} +  S^c_{-}(0)  S^a_{+}\right] \mathcal{P}_1  + J   \mathcal{P}_1S^c_{z}(0) S^a_z \mathcal{P}_1
 + U n^c(0) \\
 &= J \mathcal{P}_1 \vec{S}^c(0)\cdot \vec{S}^a  \mathcal{P}_1  + U n^c(0),
\end{align}
where 
\begin{equation}
U = \tilde{U} + \frac{J}{4} = - \frac{|V|^2}{2} \left[ \frac{1}{\epsilon_0 + U -\epsilon_F}  - 
\frac{1}{\epsilon_F - \epsilon_0}
\right]
\end{equation}
Thus, we see that virtual transitions on and off the impurity orbital with $j=\tfrac{1}{2},l=1, s=\tfrac{1}{2}$ induce
an anti-ferromagnetic interaction, which tends to anti-align the pseudo-spin of the impurity orbital and the 
conduction electrons and therefore leads to spin-flip scattering. Again, we emphasize this model is  formally identical to the Kondo model with an important difference: The pseudo-spin operators $\vec{S}^{a/c}$ do not correspond to the spin $\vec{s}$ of the electrons in the conduction band and the localized orbital but their total angular momentum $\vec{j} = \vec{l} +\vec{s}$. The virtual transitions also induce a scattering potential, $U$ ($\tilde{U}$ in Eq.~\ref{eq:hamac2}). In what follows,
we shall drop such terms and focus only on the effects of the Kondo interaction term $\propto J$. We shall return to discussing the effects
of such potential terms after we have completed our analysis of the Kondo
interaction. 

In the following section, we describe the solution of the above Kondo model within a different kind of mean field approach to that of Anderson's. 
\subsection{A Mean-field theory for the ground state of the quantum impurity in the Kondo regime} 
It was realized by a number of studies (see Ref.~\cite{HewsonKondo} for a comprehensive review) culminating with the
numerical studies of Wilson~\cite{Wilson_RevModPhys.47.773} and the formulation of the local Fermi liquid theory by 
Nozi\`eres~\cite{nozieres1974fermi,nozieres1974fermi} that the ground state of the Anderson  impurity model in the Kondo regime and its low-energy description, the Kondo model derived above, is non-magnetic. However, at the same time, the ground state exhibits some rather unusual properties. The absence of local moment in the ground state
means the local  moment predicted by Anderson's mean field approach~\cite{Anderson_PhysRev.124.41} is screened by the conduction electrons as a result of the spin-flip scattering. This is phenomenon is known as Kondo screening.

 Following the above theoretical developments,  it was pointed out by  a number of authors, but most notably by Read and Newns (see Ref.~\cite{NewnsRead_JPhysC_1983} and references therein) that the ground state properties can be captured by a mean field approach that is fundamentally different from Anderson's. For the sake of completeness and pedagogy, in this section we present an adaption of the main ideas in the work of Read and Newns~\cite{NewnsRead_JPhysC_1983} in order to illustrate the resonant nature of the scattering in the ground state of the quantum impurity model introduced above. 
 
   To make the mean-field approach mathematically rigorous requires  generalizing the model from the current SU$(N)$ spin-orbital symmetry with $N = 2$ for Ce$_x$La$_{1-x}$Cu$_6$  to an arbitrary SU$(N)$ group. Thus, we let  the indices $m,m^{\prime}$ take values in the set $1, \ldots, N$ and assuming that we are dealing with $N$ component fermions described by $c_{km}, c^{\dag}_{km}$ and $a^{\dag}_m, a_m$. In addition, in order to obtain a sensible model in the limit where $N \to +\infty$ where both $H^c_0$ and $H_{ac}$ are $O(N)$, we need to redefine the coupling $J/2 \to J/N$ so that the interaction term $H_{ac}$ remains of $O(N)$ (recall that  $\sum_m \sim O(N)$ and $\sum_{m,m^{\prime}} \sim O(N^2)$, roughly speaking).  In this limit where $N\to +\infty$, it is mathematically possible to  neglect the fluctuations of the mean field (see discussion at the end of this section, however). 
   
     As mentioned above, because spin-flip scattering with the conduction electrons will alter the orientation of the local moment, at low temperatures the latter disappears and a complex 
   many-body state emerges, where the conduction electrons form a spin-orbital singlet with the local orbital. In other words, the  impurity (pseudo-) spin is completely screened by the conduction electrons. As a result of this singlet formation, the impurity behaves as non-magnetic potential that scatters the conduction electrons at the Fermi level with a unitary phase shift. The Kondo screening cloud can be polarized by the scattering electrons, which induces an additional residual local interaction, whose effects on the scattering phase shift can be neglected at zero-temperature. In order to see how all this comes about, we begin by rewriting the interaction term in the Kondo Hamiltonian of  Eq.~\eqref{eq:hamac2} in terms of the operator
\begin{equation}
T_m = \sum_{k} c^{\dag}_{k m} a_{m} = c^{\dag}_m(0) a_m,
\end{equation}
which, after generalizing the symmetry from SU$(2)$ to SU$(N)$ yields the following expression for the Kondo interaction: 
\begin{equation}
H_{ac} = - \frac{J}{N} \sum_{k, k^{\prime},m,m^{\prime}}
a^{\dag}_{m} c_{km} c^{\dag}_{k^{\prime}m^{
\prime}} a_{m^{\prime}} =  - \frac{J}{N}\sum_{m,m^{\prime}} T^{\dag}_{m} T_{m^{\prime}}.
\label{eq:hacint}
\end{equation}
Following Read and Newns~\cite{NewnsRead_JPhysC_1983} we  assume that
the operators $T_m$ and $T^{\dag}_m$ acquire a finite expectation value, that is,  $\langle T_m\rangle = T  \neq 0$ ($\langle T^{\dag}_m\rangle = T^*\neq 0$)
where $T$ is a complex number \emph{independent} of $m$.  Thus, in the spirit of a mean field theory,
the Hamiltonian of the Kondo model can be approximated by the following mean-field Hamiltonian:
\begin{equation}
H^{MF}_{K}=  \sum_{k,m} \epsilon_k  \:  c^{\dag}_{km} c_{km} - J \sum_{k,m} \left[ T^* c^{\dag}_{km} a_{m} + T a^{\dag}_{m} c_{km} \right]. \label{eq:hammf}
\end{equation}
However, in the above treatment of $H_{ac}$ we have been rather careless of the important constraint that is
needed to define the Kondo Hamiltonian, namely that $n_a = 1$. In the discussion of the previous subsection, this was taken care of by the projector $\mathcal{P}_1$, which has been dropped at the beginning of this subsection to lighten the notation. In order to take care of this constraint properly, we shall introduce an additional Lagrange multiplier $\lambda$ and study the following Hamiltonian:
\begin{equation}
H_K(\lambda) = H_K + \lambda(n_a - Q).
\end{equation}
Next, we compute the free energy for
$H_K(\lambda)$. The latter can be obtained
from the standard expressions for the partition function (see e.g. \cite{Feynman_StatisticalMechanics}):
\begin{align}
Z\lambda) &= e^{-\beta F(\lambda)} = \mathrm{Tr} \: e^{-\beta (H_K(\lambda)-\mu  N_T)} \\
H_K(\lambda) &= H_K  + \lambda \left(n_a - Q\right), \label{eq:freenrgcon}\\
\end{align}
where $N_T = \sum_{m} a^{\dag}_m a_m + \sum_{k,m} c^{\dag}_{km} c_{km}$ 
is the total fermion number operator, $\beta = 1/k_B T$ the inverse absolute temperature,
and $\mu$ the chemical potential. It follows from \eqref{eq:freenrgcon} that by extremizing
the free energy $F(\lambda)$ with respect
to $\lambda$, i.e.
\begin{equation}
\frac{\partial F(\lambda)}{\partial \lambda }  = \langle n_a \rangle - Q = 0,
\end{equation}
we can impose the constraint on average ~\cite{NewnsRead_JPhysC_1983}. Since, in 
in practice, we shall rely on a mean-field approximation to  obtain 
$F(\lambda)$, this will prove sufficient.  To obtain the optimal mean-field approximation to the free energy,    we employ Feynman's variational principle~\cite{Feynman_StatisticalMechanics}, 
which states that the free energy  $F(\lambda)$ fulfills the following
inequality:
\begin{equation}
F(\lambda) \leq F_{V}(\lambda, T, T^*) = F_{MF}(\lambda,T,T^*) + \langle  H_{K}(\lambda) - H_{MF}(\lambda)\rangle_{MF},
\label{eq:fv}
\end{equation}
where $F_{MF}(\lambda,T,T^*)$ is the free energy of $H^{MF}_K(\lambda) = H_{K} 
+ \lambda(n_a - Q)$ and $\langle \ldots \rangle_{MF}$ stands for the thermal average
with the grand-canonical  density matrix corresponding to $H^{MF}_K(\lambda)$.
Thus, we optimize $F_V$ with respect to the all variational parameters $\lambda, T, T^*$. The resulting extremum
free energy becomes increasingly accurate as $N\to \infty$~\cite{NewnsRead_JPhysC_1983}.
The Hamiltonian $H^{MF}_K(\lambda,T,T^*)$ for the values of
the parameters that extremize $F_V$ describes the ground state of the
quantum impurity~\cite{NewnsRead_JPhysC_1983,HewsonKondo,Wilson_RevModPhys.47.773}. Computing the expectation value on the right hand side of Eq.~\eqref{eq:fv} yields:
\begin{equation}
 \langle H_{K}(\lambda) - H_{MF}(\lambda)\rangle_{MF} = J N |T|^2.
\end{equation}
In addition, the free energy for $H^{MF}_K(\lambda)$,  $F_{MF}(\lambda,T,T^*)$,  
can be obtained from the  change in the density of states, 
$\Delta \rho(\epsilon)$, which  in turn  is related to the scattering phase shift $\eta(\epsilon)$~\cite{HewsonKondo,mahan2013many}:
\begin{equation}
\Delta \rho(\epsilon) =\frac{N}{\pi} \partial_{\epsilon}\eta(\epsilon)
\end{equation}
In the wide band limit, for the resonant level model defined by $H_{K}(\lambda)$, the scattering phase shift is given
by~\cite{HewsonKondo,mahan2013many}
\begin{equation}
\eta(\epsilon) =  \tan^{-1}\left( \frac{\Gamma}{\lambda-\epsilon} \right) =  \mathrm{Im} \: \ln \left[\lambda + i\Gamma -\epsilon \right]. \label{eq:wideband}
\end{equation}
where 
\begin{equation}
\Gamma = \pi J^2 |T|^2 N_F \ll \epsilon_F
\end{equation}
is the level width ($N_F$ is the density of states at the Fermi energy), which from here on becomes one of the two variational parameters of the problem (the other one being $\lambda$). Hence, at zero temperature the free energy (i.e. the ground state energy) as a function of $\Gamma$ and $\lambda$ reads: 
\begin{align}
F_V(\lambda,\Gamma)  &=  F_0 +   \int d\epsilon \: \Delta \rho(\epsilon)\: f(\epsilon)
(\epsilon-\epsilon_F)   +   N \left( \frac{\Gamma}{\pi J \rho_0}- \lambda q\right), \label{eq:fv1}
\end{align}
where $f(\epsilon) = \theta(\epsilon_F-\epsilon)$ is the zero-temperature Fermi-Dirac distribution, $q = Q/N$, and
$F_0$ is the free (ground state) energy of the conduction electrons.  Integrating by parts the second term 
on the right hand-side of \eqref{eq:fv1} allows us to rewrite it as follows:
\begin{equation}
\int^{\epsilon_F}_{0} \frac{d\epsilon}{\pi}\: (\epsilon-\epsilon_F) \partial_{\epsilon} \eta(\epsilon) = 
\left[ \left(\epsilon-\epsilon_F\right) \frac{\eta(\epsilon)}{\pi}\right]^{\epsilon_F}_{0} - \int^{\epsilon_F}_{0} \frac{d\epsilon}{\pi} \: 
\eta(\epsilon) = -  \int^{\epsilon_F}_{0} \frac{d\epsilon}{\pi} \: \eta(\epsilon).
\end{equation}
The boundary terms vanish at $\epsilon=0$ because the phase shift at the bottom
of the band vanishes, i.e. $\eta(0) = 0$. This expression for the energy shift due to a scattering potential is generally known as Fumi's theorem~\cite{mahan2013many}.
Hence, 
\begin{equation}
F_V(\lambda, T, T^*) =  F_0  -  N \int^{\epsilon_F}_{0} \frac{d\epsilon}{\pi} \: \eta(\epsilon)
 +N \left( \frac{\Gamma}{\pi J \rho_0}- \lambda q\right). \label{eq:fvv}
\end{equation}
Looking for extrema with respect to $\lambda$ yields:
\begin{equation}
-  N \int^{\epsilon_F}_{0} \frac{d\epsilon}{\pi} \: \partial_{\lambda} \eta(\epsilon) - N q  = 
N  \int^{\epsilon_F}_{0} \frac{d\epsilon}{\pi} \: \partial_{\epsilon} \eta(\epsilon) - N q   = 
N \frac{\eta(\epsilon_F)}{\pi} - N q = 0.
\end{equation}
In this derivation, we have used that $\partial_{\lambda} \eta(\epsilon)  = -\partial_{\epsilon} \eta(\epsilon)$ (see Eq.~\ref{eq:wideband}).  This leads to the important result 
\begin{equation}
\frac{\eta(\epsilon_F)}{\pi} = q  = \frac{Q}{N} =  \frac{ \langle n_a \rangle}{N},
\end{equation}
known as Friedel's sum rule~\cite{mahan2013many,HewsonKondo}.  For this quantum impurity model, the sum rule states relates the occupation of the orbital $\langle n_a \rangle$ to the
phase shift of the conduction electrons at the Fermi level, 
$\eta(\epsilon_F)$~\cite{Langreth_PhysRev.150.516}. 
Introducing $\xi= \lambda + i \Gamma -\epsilon_F$,  Friedel's sum rule becomes:
\begin{equation}
\frac{1}{\pi} \mathrm{Im} \: \ln \left[ \xi \right]   = q, 
\end{equation}
which fixes the phase of the variable $\xi$ to be $\pi q$.
In addition, let us notice that:
\begin{equation}
\partial_{\Gamma} \eta(\epsilon) = \mathrm{Im} \frac{i}{\lambda+i\Gamma-\epsilon} =  \mathrm{Im}\:  (-i) \partial_{\epsilon}\ln[\lambda+i \Gamma - \epsilon] = - 
\partial_{\epsilon} \mathrm{Re} \: \ln[\lambda+ i \Gamma -\epsilon].
\end{equation}
Thus, the extremum condition with respect to $\Gamma$ leads
to the equation:
\begin{equation}
\partial_{\Gamma} F_V(\lambda,\Gamma) = \frac{N}{\pi} \mathrm{Re}\: \int^{\epsilon_F}_0 \partial_{\epsilon} \mathrm{Re} \: \ln[\lambda+ i \Gamma -\epsilon]) + \frac{N}{\pi N_F J} =\frac{N}{\pi} \mathrm{Re}\: \ln \left(\frac{\lambda+i\Gamma-\epsilon_F}{\lambda+i\Gamma} \right)+\frac{N}{\pi J N_F } = 0,
\end{equation}
which, in terms of $\xi$, can be rewritten as follows:
\begin{equation}
\ln \frac{|\xi|}{|\xi +\epsilon_F|} = - \frac{1}{J N_F}. 
\end{equation}
 In the limit where $J\ll \epsilon_F$,  we find
\begin{equation}
 |\xi| =  \sqrt{(\lambda-\epsilon_F)^2 + \Gamma^2} = \epsilon_F e^{-1/J N_F} = k_B T_K
 \end{equation}
The above expression defines the Kondo temperature, $T_K$.
Using  $\arg \: \xi = \pi q$, the following solutions of for mean-field parameters are obtained: 
\begin{align}
\lambda - \epsilon_F = k_B T_K \cos (\pi q),\\
\Gamma = k_B T_K \sin \left(\pi q \right).
\end{align}
Here we are interested in the cases where $N  = 2$ (doublet ground state)  and $\langle n_a \rangle = Q = 1$. 
Hence, $q = 1/2$. Using Friedel's sum rule yields $\eta(\epsilon_F) = \pi/2$ (doublet) and  $\eta(\epsilon_F) = \pi/4$ (for the quarted) for the phase shift at
the Fermi energy with $\lambda = \epsilon_F$, and $\Gamma = k_B T_K$.  Thus we conclude that, within this mean-field approach which is accurate for $N\to \infty$, the Anderson model is described by a resonant level model for which the renormalized level position and width are self-consistently determined.  When the state is a doublet ($N = 2$), the level  is \emph{pinned} exactly at the Fermi energy and has a line width equal to $k_B T_k  = \epsilon_F e^{-1/J\rho_0}$, where $J$ is the exchange coupling constant. This resonance is known as the Abrikosov-Suhl-Kondo resonance~\cite{HewsonKondo}. Note that
the resonant level is a low-energy excitation and, in that sense, it is very different from the original impurity level, which is still located at energy $\epsilon_0$ and has a width $\Delta = \pi |V|^2 N_F$ in the wide band limit. 

 Let us next briefly discuss the effects of the scattering potential terms that have been discarded above. The addition of such terms will induce an additional phase shift to the electrons, thus making the scattering shift deviate from the unitary limit where $\eta(\epsilon_F)=\pi/2$. Since we are using this model to illustrate the maximal effect of the
 anomalous velocity on the spin transport, we shall assume the effects of $U$ ($\tilde{U}$)
 are negligible, which is a good approximation deep in the Kondo regime where $\langle n_a \rangle = 1$. As discussed in the main text, in this limit, the spin Hall conductance is entirely 
 caused by the the quantum side-jump mechanism (i.e. the anomalous velocity) and skew scattering gives a vanishing contribution. 
 However, in the presence of  additional scattering channels, the contribution from skew scattering becomes non-zero. Here
 we illustrate its interplay with the unitary Kondo scattering by including an additional channel with $l=0,j=1/2,s=1/2$. See discussion in the next subsection.

 \begin{equation}
 \delta T_{m} = c^{\dag}_{m}(0) a_m - T.
 \end{equation}
 Thus, 
 \begin{equation}
H_{K} = H^c_0 - \frac{J}{N}
\sum_{m,m^{\prime}} \left( T^* + \delta T^{\dag}_m \right) \left( T+ \delta T_{m^{\prime}} \right) = H^{MF}_K + J N |T|^2 +H_{\mathrm{fl}}, 
 \end{equation}
where the fluctuation Hamiltonian reads:
\begin{equation}
H_{\mathrm{fl}} =- \frac{J}{N}\sum_{m} \delta T^{\dag}_m \delta T_m
\end{equation}
In the basis of scattering states that diagonalizes the mean-field Hamiltonian, $H^{MF}_{K}$ the states created by $a^{\dag}_m$ and $c^{\dag}_{km}$ become admixed.
This means that the fluctuation energy describes an interaction between the conduction electrons in the $j=\tfrac{1}{2}, l=1, s = \tfrac{1}{2}$ channel that takes place at the impurity position. Being a weak ($J \ll \epsilon_F$) local interaction, it can be treated perturbatively and leads to subleading corrections to he free energy of the impurity  at low temperatures~\cite{NewnsRead_JPhysC_1983}. At $T = 0$, its effect on the scattering phase-shift of the conduction electrons at the Fermi energy vanishes~\cite{Wilson_RevModPhys.47.773,HewsonKondo}. The ground state
is therefore entirely described by a non-magnetic impurity with unitary phase shift   $\delta = \frac{\pi}{2}$ as required by Friedel's sum rule, i.e. by occupation of the impurity level $\langle n_a \rangle$.  Nonetheless, the effect fluctuations is important in determining other properties of the impurity such as the so-called Wilson ratio, $W$. The latter is the ratio of the impurity contribution to the magnetic susceptibility $\chi$ and its contribution to the linear coefficient of the specific heat $\gamma$. When the effect of the fluctuations described by $H_{\mathrm{fl}}$ is taken into account to leading order in $1/N$, the Wilson ratio is found to be~\cite{NewnsRead_JPhysC_1983}:
 \begin{equation}
 W = \frac{\chi}{\gamma} = \frac{1}{1-\frac{1}{N}}.
 \end{equation}
This is an indication of the rather unusual properties of the local Fermi liquid that describes the low-temperature properties of the quantum impurity. Let us point out that, for a non-interacting Fermi liquid, $W = 1$, because both $\chi$ and $\gamma$ are determined by the density of states af the Fermi energy, $N_F$.
Therefore, deviations from unity are indications of strongly correlated behavior.  However, for $N = 2$, fluctuations effects are important and the above formula predicts that Wilson ratio strongly deviates from its non-interacting value, $W  = 2$. Interestingly, this value is in agreement with the results obtained using more sophisticated approaches~\cite{HewsonKondo,Wilson_RevModPhys.47.773}.
 \subsection{T-matrix for the Quantum Impurity model in the Kondo screened regime}
 \label{subsec:tmat}

 Our next goal is to obtain the matrix elements of the quantum impurity T-matrix, $T^{R/A}_{\vec{k}\vec{p}}$. 
Since in the presence of SOC, the total angular momentum $\vec{j} = \vec{l} + \vec{s}$ is the good quantum number,
we shall project the T-matrix in scattering channels that correspond to the multiplets of $\vec{j}$ and are therefore
labelled by quantum numbers $(l, j, m)$ ($s = 1/2$ for all of them). 

 We begin by considering the expansion of the on-shell scattering $S$ and $T$ matrices,
\begin{align}
\hat{S} (\vec{k},\vec{p}) = \delta_{\vec{k},\vec{p}} \mathbb{1} - 2 \pi i \delta(\epsilon_k - \epsilon_p) \hat{T}^R(\vec{k},\vec{p})\label{eq:smatrix} ,
\end{align}
in terms of total angular momentum projectors, $\hat{P}^{j,l}(\vec{\hat{k}},\vec{\hat{p}})$. 
Using the following identity:
\begin{equation}
    \delta^{(2)}(\vec{\hat{k}}-\vec{\hat{p}})\mathbb{1}=\sum_{l,j} \hat{P}^{j,l}(\vec{\hat{k}},\vec{\hat{p}})
\end{equation}
where $l,j$ run over all possible values, we can expand the delta-function, $S$-matrix and $T$-matrix as follows:
\begin{equation} \label{eq:delta_function}
    \delta_{\vec{k},\vec{p}} \mathbb{1} \to (2\pi)^3 \delta^{(3)}(\vec{k} - \vec{p}) \mathbb{1}= \delta(\epsilon_k-\epsilon_p) \sum_{j,l} \frac{4\pi}{N(\epsilon_k)}   \hat{P}^{j,l}  (\vec{\hat{k}},\vec{\hat{p}}),
\end{equation}
\begin{equation}
\hat{T}^R(\vec{k},\vec{p}) = \sum_{j,l} t_{j,l}(\epsilon_k) \hat{P}^{j,l}(\vec{\hat{k}},\vec{\hat{p}}),
\end{equation}
\begin{equation}
\hat{S}(\vec{k},\vec{p}) = \delta(\epsilon_k-\epsilon_p)  \sum_{j,l} \frac{4\pi s_{j,l}(\epsilon_k)}{ N(\epsilon_k)} \hat{P}^{j,l} (\vec{\hat{k}},\vec{\hat{p}}).
\end{equation}
In the second step of Eq.~\eqref{eq:delta_function}, we used $\langle \vec{k}| \vec{p} \rangle = \int d^{3}r e^{i\vec{r}\cdot (\vec{k}-\vec{p})}=(2\pi)^{3}\delta^{(3)}(\vec{k}-\vec{p})$ where, as usual, we set the volume to unity. Introducing  these expressions into Eq.~\eqref{eq:smatrix}, we arrive at the following relationship:
\begin{equation}
t^R_{j,l}(\epsilon_k) = 2 \left[\frac{1-s_{j,l}(\epsilon_k)}{i N(\epsilon_k)}\right] = - \frac{4 e^{i\eta_{j,l}(\epsilon_k)}}{N(\epsilon_k)} \sin\eta_{j,l}(\epsilon_k).
\end{equation}
In deriving the above, we have used that the (unitary) $S$-matrix is diagonal within each multiplet, that is,  
\begin{equation}
s_{j,l}(\epsilon_k) = e^{2i\eta_{j,l}(\epsilon_k)},\label{eq:seig}
\end{equation}
where $\eta_{j,l}(\epsilon_k)$ is the scattering phase shift for the multiplet $(j,l)$. 

 In the simplified quantum impurity model considered in the previous subsection, we relevant scattering channels
 are $l = 0, j=1/2$ and $l=1, j =1/2$ and $j=3/2$. The  projectors $\hat{P}^{l,j}$ can be obtained from \eqref{eq:c11}
 and \eqref{eq:c12} for $(l=1,j=1/2)$, and  \eqref{eq:c01} together with \eqref{eq:c02} for  $(l=0,j=1/2)$
 by considering:
 \begin{align}
 \sum_{m=\pm} |k m\rangle \langle k m | &= \sum_{\alpha,\beta}\int \frac{d\vec{\hat{k}}d\vec{\hat{p}}}{(4\pi)^2} \, 
  \hat{P}^{j=1/2,l=1}_{\alpha\beta}(\vec{\hat{k}},\vec{\hat{p}})
 | \vec{k} \alpha \rangle \langle \vec{p} \beta |, \\
 \sum_{m=\pm} |0 k  m\rangle \langle 0 k  m | &= \sum_{\alpha,\beta}\int \frac{d\vec{\hat{k}}d\vec{\hat{p}}}{(4\pi)^2} \, 
  \hat{P}^{j=1/2,l=0}_{\alpha\beta}(\vec{\hat{k}},\vec{\hat{p}})
 | \vec{k} \alpha \rangle \langle \vec{p} \beta |,\\
  \sum_{m=-3/2}^{+3/2} |k m\rangle \langle k m | &= \sum_{\alpha,\beta}\int \frac{d\vec{\hat{k}}d\vec{\hat{p}}}{(4\pi)^2} \, 
  \hat{P}^{j=3/2,l=1}_{\alpha\beta}(\vec{\hat{k}},\vec{\hat{p}})
 | \vec{k} \alpha \rangle \langle \vec{p} \beta |,
 \end{align}
 where $|km\rangle = c^{\dag}_{km}|0\rangle$ ($l=1$) and $|0km \rangle = c^{\dag}_{0km}|0\rangle$ ($l=0$).
 In the above expressions, the projectors are
\begin{align}
\hat{P}^{j=1/2,l=1}_{\alpha\beta}(\vec{\hat{k}}, \vec{\hat{p}}) &= \sum_{m=\pm} F^{l=1,j=1/2}_{\alpha m}(\vec{\hat{k}})
\left[ F^{l=1,j=1/2}_{\beta m}(\vec{\hat{p}})\right]^*  
= \frac{1}{4\pi}\left[ (\vec{\hat{k}}\cdot \vec{\hat{p}}) \delta_{\alpha\beta} + i 
(\vec{\hat{k}}\times \vec{\hat{p}})\cdot \vec{\sigma} )_{\alpha\beta} \right], \\
\hat{P}^{j=1/2,l=0}_{\alpha\beta}(\vec{\hat{k}},\vec{\hat{p}}) &=  
\sum_{m=\pm} F^{l=0,j=1/2}_{\alpha m}(\vec{\hat{k}})
\left[ F^{l=0,j=1/2}_{\beta m}(\vec{\hat{p}})\right]^* =
\frac{1}{4\pi}\delta_{\alpha\beta},\\
\hat{P}^{j=3/2,l=0}_{\alpha\beta}(\vec{\hat{k}},\vec{\hat{p}}) &=  
\sum_{m=\pm} F^{l=0,j=3/2}_{\alpha m}(\vec{\hat{k}})
\left[ F^{l=0,j=3/2}_{\beta m}(\vec{\hat{p}})\right]^* =\frac{1}{4\pi}\left[ 2 (\vec{\hat{k}}\cdot \vec{\hat{p}}) \delta_{\alpha\beta} - i 
(\vec{\hat{k}}\times \vec{\hat{p}})\cdot \vec{\sigma} )_{\alpha\beta} \right]
\end{align}

Retaining only these three channels, the (on-shell) T-matrix for electrons at the Fermi energy takes the following form:
\begin{align} \label{eq:t-matrix_final}
\hat{T}^{R}(\vec{k},\vec{p}) =-\frac{e^{i\eta_{0}} \sin \eta_{0}}{\pi  N_F} \mathbb{1}
 - \frac{e^{i\eta_{1}}}{\pi k^2_F N_F}  \sin\eta_{1} \left[  (\vec{k}\cdot \vec{p})  + i (\vec{k}\times\vec{p})\cdot \vec{\sigma}\right] - 
 \frac{e^{i\eta_{2}}}{\pi k^2_F N_F}  \sin\eta_{2} \left[ 2 (\vec{k}\cdot \vec{p})  - i (\vec{k}\times\vec{p})\cdot \vec{\sigma}\right]
\end{align}
where we have introduced the following short-hand  notation for the phase-shifts at the Fermi energy:
\begin{equation}
    \eta_{0} \equiv
    \eta_{j=\tfrac{1}{2},l=0}(\epsilon_F),
    \quad
    \eta_{1} \equiv \eta_{j=\tfrac{1}{2},l=1}(\epsilon_F),
  \quad 
  \eta_{2} \equiv \eta_{j=\tfrac{3}{2},l=1}(\epsilon_F).
\end{equation}
For the channels for which electron correlations (i.e. Kondo screening)  do not play a role, 
 the level position and width do not need to be   determined  self-consistently as we have done in the previous section using the large $N$ mean-field approach. In principle,  they determined by parameters that determine the impurity potential. Thus, for instance,
\begin{equation}
\eta_0\equiv  \eta_{j=\frac{1}{2}, l = 0} (\epsilon_F) = \tan^{-1}\left(\frac{\tilde{\Delta}}{\tilde{\epsilon}_0-\epsilon_F} \right).   
\end{equation}
where we have used the wide-band approximation again and therefore $\tilde{\Delta} = \pi |\tilde{V}|^2 N_F$. 
In practice, we $|eta_0|$ is treated as small and taken as a fitting parameter $\eta_0 \sim \pm 0.1$~\cite{Fert_resonant1,Fert_resonant2,Fert_resonant3}. On the other hand, the phase shifts for the doublet ($l=1,j=1/2$) is determined the strong electronic correlation which results in  Kondo screening. As mentioned above,  for the 
Ce$_x$La$_{1-x}$Cu$_6$ alloy, the doublet is the ground state with $\eta_1 = \pi/2$ as corresponds to $N=2$, and  we shall treat the quartet as weakly coupled with $|\eta_2| \ll 1$.  This concludes the derivation of the single-impurity T-matrix for the quantum impurity model describing Ce in these alloys. 

 Up to this point, our discussion has been  concerned  with the T-matrix on the energy shell, that is, for $|\vec{k}| = |\vec{p}|$,
 and energy $\epsilon = \epsilon_{\vec{k}} = \epsilon_{\vec{p}}$. However, in order to compute the
 anomalous velocity resulting from the gradient expansion of the collision integral, we need to 
evaluate derivatives of the T-matrix off the energy shell. To this end, we can take a step back
from the above considerations and use the following expression for the (off-shell) 
T-matrix~\cite{HewsonKondo}:
\begin{equation}
T^R_{\vec{k}\vec{p}}(\epsilon) = V^*_{\vec{k}} G^R_{aa}(\epsilon) V_{\vec{p}}, 
\end{equation}
where $V_{\vec{p}}$ is the hybridization matrix element with the orbital, 
and $G^R_{aa}(\epsilon)$ the orbital Green's function. This result applies 
both to interacting and non-interacting quantum impurities~\cite{HewsonKondo}. Using this expression, we see that
the derivatives with respect to $\vec{k}$ and $\vec{p}$ affect
the hybridization matrix elements $V^*_{\vec{k}}$ and $V_{\vec{p}}$.
Notice that the rotational symmetry of the orbital dictates that
\begin{equation}
V_{\vec{p}}  = V_{p} F^{l,j}(\vec{\hat{p}}). 
\label{eq:vp1}
\end{equation}
where $V_p = V_{|\vec{p}|}$. In addition, for $l=1$, the momentum dependence of the radial part of the scattering states implies that $V_{p} = p R(p)$, where $R(p) \simeq R_0 + R_1 p^2$ with $p^2_F R_1  \ll R_0$~\footnote{We have explicitly verified this for our model by explicitly calculating the momentum dependence of $V_p$ as the matrix element of the impurity Coulomb potential between a plane wave and a  $p$ orbital that is properly orthogonalized to the conduction band states.}
\begin{equation}
V_{\vec{p}}  = R_{p} F^{l=1,j}(\vec{p}). 
\end{equation}
In the above expression (compared to \eqref{eq:vp1}) the leading linear behavior of $V_p$ with $p$ has been  absorbed into $F^{l=1,j}(\vec{p})$ by replacing the unit
vector $\vec{\hat{p}}$ by the full vector $\vec{p}$. Hence, 
\begin{equation}
\vec{\nabla}_{\vec{p}} V_{\vec{p}} =  F^{(l=1,j)}(\vec{p})\:  \left( \frac{dR_p}{dp}  \frac{\vec{p}}{p}\right) + 
R_{p} \vec{\nabla}_{\vec{p}} F^{l=1,j}(\vec{p}) \simeq \left(\frac{V_p}{p} \right)     \vec{\nabla}_{\vec{p}} F^{l=1,j}(\vec{p})
\end{equation}
Note that,  if we take the derivative of the on-shell T-matrix the contribution of the first
term proportional  to $dR_p/dp$ in the expression would be missed. 
Instead we get an (incorrect) 
additional contribution from the derivative of the energy dependence of the phase shift since $\epsilon=\epsilon_k = \epsilon_p$.  However,
as we have explained above $G_p$ is a slowly varying function of $p$ and its derivative with respect to $p$ can be safely neglected. Thus, 
\begin{equation}
\vec{\nabla}_{\vec{p}} V_{\vec{p}}  \simeq \left(\frac{V_p}{p} \right)     \vec{\nabla}_{\vec{p}} F^{l=1,j}(\vec{p})
\end{equation}
Thus, it is a good approximation to apply the derivatives to the on-shell
T-matrix provided the latter only act upon the angular part described by the projectors 
$P^{j,l}(\vec{\hat{k}},\vec{\hat{p}})$ and not on the energy dependence of the phase-shifts.

\section{On the positivity of the relaxation matrix}\label{sec:positivity}

In Sec.~\ref{sec:derivation}, we have introduced the following relaxation super-operator:
\begin{equation}
\Lambda_{\alpha\beta,\gamma\delta}(\vec{p},\vec{k})=\delta_{\vec{p}\vec{k}}\delta_{\alpha\gamma}\delta_{\beta\delta}-S_{\alpha\gamma}(\vec{p},\vec{k})S_{\beta\delta}^{*}(\vec{p},\vec{k}),
\end{equation}
This operator describes the relaxation of small deviations from equilibrium
$\delta n_{\alpha\beta}$ according to \eqref{eq:relaxation}. Using the unitarity of the $S$-matrix, 
the right hand-side of Eq.~\eqref{eq:relaxation} be recast (again) as Lindbladian:
\begin{equation}
\partial_t\delta n_{\vec{p}}(t) = -\frac{n_{im}}{2\pi} \sum_{\vec{k}} \left\{ 
\frac{1}{2} \left[S(\vec{p},\vec{k}) S^{\dag}(\vec{k},\vec{p})  
+ S^{\dag}(\vec{p},\vec{k}) S(\vec{k},\vec{p})\right] 
\delta n_{\vec{p}}(t) - S(\vec{p},\vec{k}) \delta n_{\vec{k}}(t) S^{\dag}(\vec{k},\vec{p})
\right\} 
\end{equation}
Linbladians are completely positive evolution superoperators that respect unitarity (see e.g. Ref.~\cite{breuer2002theory}). However, many of the results on Lindbladians are not entirely relevant here because the above density matrices $\delta n_{\vec{p}}$ are not normalized in the same way as ordinary density matrices in quantum mechanics. Recall that
\begin{equation}
\sum_{\vec{p}}\mathrm{Tr} \,  n_{\vec{p}} = 2 \sum_{\vec{p}} n_F(\epsilon_{p}) = N,
\end{equation}
where  $n_F(\epsilon)$ is the Fermi-Dirac distribution and $N$ the total number of electrons in the system. This  implies that $\sum_{\vec{p}} \delta(\epsilon_p-\epsilon_F) \mathrm{Tr}\, 
\delta  n_{\vec{p}} = 0$ for a uniform system. Thus, we have devised a proof of the positivity of the superoperator $\Lambda$, which is described below.

In order to define positivity, let  us first  define the scalar product of two distribution density matrices $\delta n$ and $\delta n^{\prime}$ as follows:
\begin{equation}
\langle \delta n |\delta n^{\prime} \rangle = \sum_{\vec{p}} \delta(\epsilon_{p}-\epsilon_F)   \mathrm{Tr}\: \left( \delta n_{\vec{p}}\delta n^{\prime}_{\vec{p}} \right).
\end{equation}
Let  us choose $\delta n^{\prime}_{\vec{p}}$
as the distribution that results from acting with the relaxation matrix upon $\delta n_{\vec{p}}$, i.e.
\begin{equation}
\delta n^{\prime}_{\vec{p}} = \delta n_{\vec{p}} -  S(\vec{p},\vec{k}) \delta n(\vec{k}) S^{\dag}(\vec{k},\vec{p})     
\end{equation}
Hence, 
\begin{equation}
\langle \delta n |\delta n^{\prime} \rangle = \sum_{\vec{p}} \delta(\epsilon_p-\epsilon_F) \left[ \mathrm{Tr} \: \delta n^2_{\vec{p}} - \sum_{\vec{k}}  \mathrm{Tr} \: \delta n_{\vec{p}} S(\vec{p},\vec{k}) \delta n_{\vec{k}} S^{\dag}(\vec{k},\vec{p}) \right] \equiv \langle \delta n | \Lambda \delta n\rangle . 
\end{equation}
Next, instead of the cases of plane waves with particular spin orientation, we shall compute the traces in the above expression in the basis that renders 
the S-matrix diagonal. 
For instance,  in the case of rotational invariant systems, this basis  are the states from the multiplets of the total angular momentum $\vec{j}$. Mathematically, 
\begin{equation}
S = \sum_{j} e^{i \eta_j} | j \rangle \langle j |
\end{equation}
where $\eta_j$ are the phase-shifts and $|j\rangle$ the S-matrix
eigen vectors. Thus,
\begin{align}
\langle \delta n | \Lambda \delta n\rangle  = \sum_{j,j^{\prime}} \langle j| \delta n |j^{\prime} \rangle \langle j^{\prime} | \delta n | j\rangle 
\left[ 1- e^{i \left(\eta_{j}-\eta_{j^{\prime}}\right)} \right]\\
= \sum_{j,j^{\prime}} |\langle j| \delta n |j^{\prime} \rangle|^2 \left[1 - \cos\left(\eta_j - \eta_{j^{\prime}}\right) \right] \geq 0 
\end{align}
For a general distribution and a non-trivial
scattering matrix with at least one $\eta_j \neq 0$, 
we expect the inequality to hold. Thus, the relaxation 
matrix is positive definite, which means that
\begin{equation}
\frac{1}{2} \partial_t \langle  \delta n(t)|  \delta n(t) \rangle = \langle \delta n(t) | \partial_{t}\delta n(t)  \rangle =  - \frac{n_{\mathrm{im}}}{2\pi}
\langle  \delta n(t) | \Lambda 
\delta n(t) \rangle < 0.
\end{equation}
%

\bibliographystyle{ieeetr}
\bibliography{ref.bib}

\end{document}